%% file: main.tex
\documentclass[twocolumn]{aastex631}

\newcommand{\water}{H$_2$O}
\newcommand{\coo}{CO$_2$}
\newcommand{\isocoo}{$^{13}$CO$_2$}
\newcommand{\ocn}{OCN$^-$}
\newcommand{\methenol}{CH$_3$OH}

\newcommand{\lbol}{$L_{\mathrm{bol}}$}
\newcommand{\lsun}{$L_{\mathrm{\odot}}$}
\newcommand{\tbol}{$T_{\mathrm{bol}}$}

\submitjournal{ApJ}

\shorttitle{Spatial Distribution of Ices in HOPS 370}
\shortauthors{Tyagi et al.}
\usepackage{rotating}
\usepackage{graphicx}
\usepackage{amsmath}
\usepackage{tikz}

\begin{document}

\title{JWST-IPA: Chemical Inventory and Spatial Mapping of Ices in the Protostar HOPS370 -- Evidence for an Opacity Hole and Thermal Processing of Ices}

\input{AuthorList}

%% Mark off the abstract in the ``abstract'' environment. 
\begin{abstract}
The composition of protoplanetary disks, and hence the initial conditions of planet formation, may be strongly influenced by the infall and thermal processing of material during the protostellar phase.  
Composition of dust and ice in protostellar envelopes, shaped by energetic processes driven by the protostar, serves as the fundamental building material for planets and complex organic molecules.
As part of the JWST GO program, ``Investigating Protostellar Accretion" (IPA), we observed an intermediate-mass protostar HOPS 370 (OMC2-FIR3) using NIRSpec/IFU and MIRI/MRS. This study presents the gas and ice phase chemical inventory revealed with the JWST in the spectral range of $\sim$2.9 to 28 \micron\ and explores the spatial variation of volatile ice species in the protostellar envelope. We find evidence for thermal processing of ice species throughout the inner envelope.
We present the first high-spatial resolution ($\sim 80$ au) maps of key volatile ice species \water, \coo, \isocoo, CO, and \ocn, which reveal a highly structured and inhomogeneous density distribution of the protostellar envelope, with a deficiency of ice column density that coincides with the jet/outflow shocked knots. 
Further, we observe high relative crystallinity of \water\ ice around the shocked knot seen in the
H$_2$ and OH wind/outflow, which can be explained by a lack of outer colder material in the envelope along the line of sight due to the irregular structure of the envelope. These observations show clear evidence of thermal processing of the ices in the inner envelope, close to the outflow cavity walls, heated by the luminous protostar.

\end{abstract}

%% https://astrothesaurus.org

%% From the front matter, we move on to the body of the paper.
%% Sections are demarcated by \section and \subsection, respectively.
%% Observe the use of the LaTeX \label
%% command after the \subsection to give a symbolic KEY to the
%% subsection for cross-referencing in a \ref command.
%% You can use LaTeX's \ref and \label commands to keep track of
%% cross-references to sections, equations, tables, and figures.
%% That way, if you change the order of any elements, LaTeX will
%% automatically renumber them.
%%
%% We recommend that authors also use the natbib \citep
%% and \citet commands to identify citations.  The citations are
%% tied to the reference list via symbolic KEYs. The KEY corresponds
%% to the KEY in the \bibitem in the reference list below. 

\input{S1_introduction}

\input{S2_ObsDataReduction}

\input{S3_Results}

\input{S4_Discussion}

\input{S5_Concluison}

\input{S6_7_Acknowledgment}

\facility{JWST (NIRSpec, MIRI)}

\software{Astropy \citep{astropy:2013, astropy:2018, astropy:2022}; CARTA \citep{angus_comrie_2021_4905459}; Matplotlib \citep{matplotlibHunter:2007}; NumPy \citep{numpyharris2020array}}; SciPy \citep{scipy2020SciPy-NMeth}.

\bibliography{main}
\bibliographystyle{aa_url}

\input{S8_Appendix}

\end{document}

%% file: AuthorList.tex
\author[0000-0002-9497-8856]{Himanshu Tyagi}
\affiliation{Department of Astronomy and Astrophysics Tata Institute of Fundamental Research \\Homi Bhabha Road, Colaba, Mumbai 400005, India}

\correspondingauthor{Himanshu Tyagi}
\email{himanshu.tyagi@tifr.res.in, tyagihimanshu027@gmail.com}

\author[0000-0002-3530-304X]{P. Manoj}
\affiliation{Department of Astronomy and Astrophysics Tata Institute of Fundamental Research \\Homi Bhabha Road, Colaba, Mumbai 400005, India}

\author[0000-0002-0554-1151]{Mayank Narang}
\affiliation{Academia Sinica Institute of Astronomy \& Astrophysics, \\ 11F of Astro-Math Bldg., No.1, Sec. 4, Roosevelt Rd., Taipei 10617, Taiwan}
\affiliation{Department of Astronomy and Astrophysics Tata Institute of Fundamental Research \\Homi Bhabha Road, Colaba, Mumbai 400005, India}

\author[0000-0001-7629-3573]{S. Thomas Megeath}
\affiliation{University of Toledo, Toledo, OH, US}

\author[0000-0001-6144-4113]{Will R. M. Rocha}
\affiliation{Laboratory for Astrophysics, Leiden Observatory, Universiteit Leiden, Leiden, Zuid-Holland, NL}

\author[0000-0001-7826-7934]{Nashanty Brunken}
\affiliation{Leiden Observatory, Universiteit Leiden, Leiden, Zuid-Holland, NL}

\author[0000-0001-8790-9484]{Adam E. Rubinstein}
\affiliation{University of Rochester, Rochester, NY, US}

\author[0000-0002-6447-899X]{Robert Gutermuth}
\affiliation{University of Massachusetts Amherst, Amherst, MA, US}

\author[0000-0001-5175-1777]{Neal J. Evans II}
\affiliation{Department of Astronomy, The University of Texas at Austin, 2515 Speedway, Stop C1400, Austin, Texas 78712-1205, USA}

\author[0000-0001-7591-1907]{Ewine F. Van Dishoeck}
\affiliation{Leiden Observatory, Universiteit Leiden, Leiden, Zuid-Holland, NL}
\affiliation{Max-Planck Institut f\"ur Extraterrestrische Physik, Garching bei München, DE}

\author[0000-0002-6136-5578]{Samuel Federman}
\affiliation{University of Toledo, Toledo, OH, US}

\author[0000-0001-8302-0530]{Dan M. Watson}
\affiliation{University of Rochester, Rochester, NY, US}

\author[0000-0001-8341-1646]{David A. Neufeld}
\affiliation{William H. Miller III Department of Physics and Astronomy, The Johns Hopkins University, Baltimore, MD, USA}

%#
\author[0000-0002-7506-5429]{Guillem Anglada}
\affiliation{Instituto de Astrof{\'i}sica de Andaluc{\'i}a, CSIC, Glorieta de la
Astronom{\'i}a s/n, E-18008 Granada, ES}

\author[0000-0002-1700-090X]{Henrik Beuther}
\affiliation{Max Planck Institute for Astronomy, Heidelberg, Baden Wuerttemberg, DE}

\author[0000-0001-8876-6614]{Alessio Caratti o Garatti}
\affiliation{INAF-Osservatorio Astronomico di Capodimonte, IT}

\author[0000-0002-4540-6587]{Leslie W. Looney}
\affiliation{Department of Astronomy, University of Illinois, 1002 West Green St, Urbana, IL 61801, USA}
\affil{National Radio Astronomy Observatory, 520 Edgemont Rd., Charlottesville, VA 22903 USA} 

\author[0000-0002-4448-3871]{Pooneh Nazari}
\affiliation{Leiden Observatory, Universiteit Leiden, Leiden, Zuid-Holland, NL}

\author[0000-0002-6737-5267]{Mayra Osorio}
\affiliation{Instituto de Astrof{\'i}sica de Andaluc{\'i}a, CSIC, Glorieta de la Astronom{\'i}a s/n, E-18008 Granada, ES}

\author[0000-0002-5812-9232]{Thomas Stanke}
\affiliation{Max-Planck Institut f\"ur Extraterrestrische Physik, Garching bei München, DE}

\author[0000-0001-8227-2816]{Yao-Lun Yang}
\affiliation{RIKEN Cluster for Pioneering Research, Wako-shi, Saitama, 351-0106, Japan}

%#
\author[0000-0001-7491-0048]{Tyler L. Bourke}
\affiliation{SKA Observatory, Jodrell Bank, Lower Withington, Macclesfield SK11 9FT, UK}

\author[0000-0002-3747-2496]{William J. Fischer}
\affiliation{Space Telescope Science Institute, 3700 San Martin Drive, Baltimore, MD 21218, US}

\author[0000-0001-9800-6248]{Elise Furlan}
\affiliation{Caltech/IPAC, Pasadena, CA, US}

\author[0000-0003-1665-5709]{Joel Green}
\affiliation{Space Telescope Science Institute, 3700 San Martin Drive, Baltimore, MD 21218, US}

\author[0000-0002-2667-1676]{Nolan Habel}
\affiliation{Jet Propulsion Laboratory, Pasadena, CA, US}

\author[0000-0001-9443-0463]{Pamela Klaassen}
\affiliation{United Kingdom Astronomy Technology Centre, Edinburgh, GB}

\author[0000-0003-3682-854X]{Nicole Karnath}
\affiliation{Space Science Institute, Boulder, CO, US}
\affiliation{Center for Astrophysics Harvard \& Smithsonian, Cambridge, MA, US}

\author[0000-0002-8115-8437]{Hendrik Linz}
\affiliation{Max Planck Institute for Astronomy, Heidelberg, Baden Wuerttemberg, DE}
\affiliation{Friedrich-Schiller-Universit\"at, Jena, Th\"uringen, DE}

\author[0000-0002-5943-1222]{James Muzerolle}
\affiliation{Space Telescope Science Institute, Baltimore, MD, US}

\author[0000-0002-6195-0152]{John J. Tobin}
\affil{National Radio Astronomy Observatory, 520 Edgemont Rd., Charlottesville, VA 22903 USA} 

\author[0000-0002-4026-126X]{Prabhani Atnagulov}
\affiliation{Ritter Astrophysical Research Center, Dept. of Physics and Astronomy, University of Toledo, Toledo, OH, US}

\author[0000-0002-5350-0282]{Rohan Rahatgaonkar}
\affiliation{Gemini South Observatory, La Serena, CL}

\author[0000-0002-9209-8708]{Patrick Sheehan}
\affil{National Radio Astronomy Observatory, 520 Edgemont Rd., Charlottesville, VA 22903 USA} 

\author[0000-0002-7433-1035]{Katerina Slavicinska}
\affiliation{Leiden Observatory, Universiteit Leiden, Leiden, Zuid-Holland, NL}

\author[0000-0003-2300-8200]{Amelia M.\ Stutz}
\affiliation{Departamento de Astronom\'{i}a, Universidad de Concepci\'{o}n,Casilla 160-C, Concepci\'{o}n, Chile}

\author[0000-0002-9470-2358]{Lukasz Tychoniec}
\affiliation{European Southern Observatory,
Garching bei M\"unchen, DE}

\author[0000-0002-0826-9261]{Scott Wolk}
\affiliation{Center for Astrophysics Harvard \& Smithsonian, Cambridge, MA, US}

%% file: S1_introduction.tex
\section{Introduction}
\label{intro}
Planetary systems are the natural byproducts of the star formation process \citep{Hartmann2009apsf.book.....H, StahlerBook}. Star formation begins with the collapse of a slowly rotating molecular cloud, followed by the formation of a central protostar and a circumstellar disk due to angular momentum conservation \citep{Terebey1984ApJ...286..529T, Hartmann2009apsf.book.....H}. The circumstellar disk serves as the birthplace of planets and is widely known as protoplanetary disk \citep{Armitage2020apfs.book.....A}. 
Terrestrial planets, such as Earth, form in the inner disk regions, where temperatures are relatively high, resulting in planetesimals relatively depleted in volatiles.
% volatile depleted planetesimals. 
Yet, the Earth is enriched in volatiles relative to planetesimals in the inner disk\citep{2012morbidelli, vanDishoeck2014prpl.conf..835V, Oberg2021}.

How did the Earth get its volatile materials? 
% started a debate on the \textit{Reset vs. Inheritance}. 
Several studies \citep[e.g.,][]{2012morbidelli, 2014Pontoppidan, Hartogh2011Natur.478..218H, Altwegg2019ARA&A..57..113A} suggest that comets and other icy bodies provided the Earth with much of its complement of volatiles. It is thought that comets are the most pristine objects in the solar system, and their chemical composition has not been altered since the solar protoplanetary disk stage \citep{Mumma2011, 2014Pontoppidan}. This begs the question, is the chemical makeup of comets and the volatiles on Earth protoplanetary or interstellar in origin?

The debate over the origin centers around two main hypotheses: 1. chemical inheritance and evolution and 2. reset. 
Chemical inheritance refers to the idea that chemical composition is inherited from the molecular cloud from which the protoplanetary disk is formed and should therefore be similar to the composition of the molecular cloud \citep{Cleeves2014Sci...345.1590C}.  In contrast, the reset hypothesis posits that the interstellar chemical composition is destroyed/reset and that the chemical inventory depends only on the chemical reactions taking place in the protoplanetary disks during the later stages of planet formation. The idea of reset is supported by meteoritic evidence (see for details \citealt{Mumma2011}; \citealt{2014Pontoppidan}). Moreover, several studies in the literature suggest contributions of both inheritance and reset \citep{2006Brownlee, Visser2009A&A...495..881V, Ciesla2010Icar..208..455C, 2018Rubin, 2019Drozdovskaya}, but the relative importance of inheritance vs reset is poorly constrained \citep{Oberg2021}. 
However, the problem might not be dichotomous as intermediate stages of star formation, particularly the protostellar phase, undergo energetic events that might trigger chemical reactions to change the chemical makeup of the natal envelope.

The current epoch of star and planet formation within our Galaxy serves as a natural laboratory to understand the origin and formation mechanisms of our solar system and all planets more generally. Most importantly, the earliest phases of star formation can provide insights into the chemical evolution of the seed material for the stars and planets. The protostellar phase, often referred to as the primary accretion phase \citep{Fischer2017, Mayank2023JApA...44...92N, Federman2023ApJ...944...49F, Tobin2024arXiv240315550T}, plays a vital role in accumulating/transporting gas and icy dust from the molecular cloud onto the star and the protostellar/protoplanetary disk for planet formation \cite[the birth-place of planets; see also][]{Manara2018A&A...618L...3M, Tychoniec2020A&A...640A..19T}. During this stage, most of the stellar mass is accreted \citep{Fischer2017}, and the initial conditions for planet formation are set. 
Hints of the early onset of planet formation, such as radial substructures detected by ALMA in protostellar disks, have also been seen during the protostellar phase \citep[see e.g.,][]{ALMAPartnership2015ApJ...808L...3A, Sheehan2018, Sheehan2020, Segura-Cox2020Natur.586..228S, Ohashi2023ApJ...951....8O}. Further, simulations have shown that protostellar disks are more plausible sites for planetesimal formation than 
 protoplanetary (Class II) disks because of their large gas and dust reservoirs (\citealt{Tsukamoto2017ApJ...838..151T}; also see, \citealt{Tanaka2019MNRAS.484.1574T}).

While the protostellar phase plays a pivotal role in accretion, its impact on altering the chemical composition of dust and ice remains underexplored. 
Energetic processes occurring during this phase, such as luminosity of the protostar itself, particularly during accretion outbursts, UV emission from accretion and outflow shocks, shocks driven by jets/outflows, and X-rays from coronal activity, have the potential to significantly modify the physical structure and chemical composition of ice species within the natal envelope \citep[e.g.,][]{Arce2008ApJ...681L..21A, 2012Kim, Visser2015, 2022Kim}. Further, frequent accretion bursts can trigger and accelerate the formation reactions of complex organic molecules (COMs) in the envelope \citep{Taquet2016}. COMs have also been detected in the jet/outflow shocks from protostars \citep{Arce2008ApJ...681L..21A,  Codella2015MNRAS.449L..11C, Lefloch2017MNRAS.469L..73L}. For a detailed discussion on chemistry during the protostellar phase, the reader is referred to \citealt{Jorgensen2020ARA&A..58..727J, Oberg2021} and references therein. Therefore, it is important to understand the impact of these energetic processes in the protostellar phase in shaping the system's chemistry.

The launch of the James Webb Space Telescope (JWST) has opened up the infrared spectral window rich in the spectral features of key volatile carriers, such as water (\water), carbon dioxide (\coo), and carbon monoxide (CO), with a superior sensitivity and angular resolution compared to \textit{ISO} and \textit{Spitzer}. In particular, the Integral Field Units of Near InfraRed Spectrograph \citep[NIRSpec;][]{Boker2022, Jakobsen2022} and Mid-Infrared Instrument \citep[MIRI;][]{Rieke2015PASP..127..584R, Wright2015PASP..127..595W} offer a unique opportunity not only to investigate the chemical composition of a protostellar envelope but also to observe the spatial variations in its distribution. This technological advancement has paved the way for a spatial investigation of protostellar processes that could shape/change the chemistry of volatiles within the envelope.

Investigating Protostellar Accretion (IPA) across the mass spectrum is a JWST Cycle 1 medium GO program \citep[PID: 1802; P.I.: S Thomas Megeath;][; Watson et al. in prep]{ipa_pro, Federman2023arXiv, Mayank2024, Adam2023arXiv, Neufeld2024arXiv240407299N} to observe five protostars in a broad luminosity range (\lbol$\sim 0.2 - 10^4$ \lsun). IPA uses NIRSpec/IFU and MIRI/MRS in $2\times2$~mosaicing mode to obtain spectral cubes with a field of view of $6'' \times 6''$. The program aims to study protostellar accretion, jets/outflows \citep{Federman2023arXiv, Mayank2024}, and to characterize the chemical properties of the protostellar environment(\citealt{Brunken2024, Nazari2024arXiv240107901N, Slavicinska2024arXiv240415399S}). 
Other programs with JWST have also focused on the icy chemistry \citep{2022Yang, 2023McClure, Rocha2024A&A...683A.124R} and mass accretion/ejection processes of the protostars \citep{Beuther2023A&A...673A.121B, Francis2024A&A...683A.249F, Tychoniec2024arXiv240204343T}.

In this study, we investigate the spatial distribution of ice species and evidence for thermal processing within the envelope of the intermediate-mass protostar HOPS 370 (aka OMC2-FIR3), which is the second most luminous source in the IPA sample. It is a Class 0/I intermediate-luminosity protostar with a bolometric luminosity (\lbol) of $315.7$ \lsun~(total luminosity $\sim511$ \lsun) and a bolometric temperature (\tbol) of $71.5$ K \citep{Furlan2016, Tobin2020_LbolApJ...890..130T}. 
The central protostar is deeply embedded in the envelope. It is located at a distance of 390 pc in the northern part of the Integral Shaped Filament (ISF) in the OMC-2 region with close to an edge-on geometry \citep[inclination angle $\sim71^{\circ}$;][]{Federman2023ApJ...944...49F, Tobin2020_LbolApJ...890..130T, Tobin2020ApJ...905..162T}. Multi-wavelength observations have revealed the presence of a strong bipolar jet/outflow \citep[see e.g.,][]{Federman2023arXiv, Osorio2017ApJ...840...36O, Gonzalez2016A&A...596A..26G, Takahashi2008ApJ...688..344T}. Our JWST observations have revealed the presence of a shocked knot in the northern part of the protostellar jet \citep[see][]{Federman2023arXiv, Neufeld2024arXiv240407299N}. 
% \textcolor{red}{HOPS 370 has no sign of variability. Is there any reference that I can cite? Check Wafa.}

In this paper, we present the chemical inventory of gas and ice species and map the spatial distribution of volatile ice species present in the envelope of HOPS 370. We also search for evidence of thermal processing of ice species in the protostellar envelope.
In section \ref{obs}, we discuss the observation and data reduction. In section \ref{results}, we present the observational results for HOPS 370. In section \ref{discussion}, we delve into an in-depth discussion of the findings and their implications and summarize our findings in section \ref{conclusion}.

%% file: S2_ObsDataReduction.tex
\begin{figure*}
    \centering
    \includegraphics[width=\linewidth]{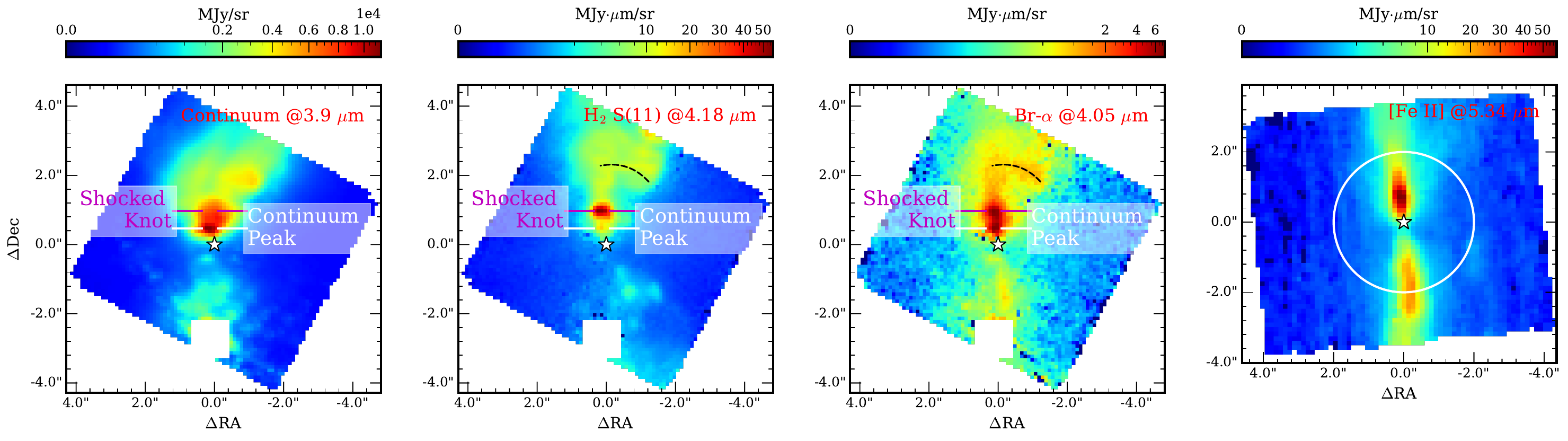}
    \caption{Morphology of the HOPS 370 protostellar environment seen in the continuum map at 3.9~\micron, H$_2$ S(11) map at 4.18~\micron, Br-$\alpha$ map at 4.05~\micron, and [Fe II] map at 5.34~\micron. All the line maps are continuum subtracted. The white star marks the ALMA continuum position in all the panels. The white circular aperture in the right panel is used for 1D spectrum extraction. The locations of the continuum peak and the shocked knot are shown using the horizontal white and magenta color lines, respectively. The dashed black line in the middle panels marks the fish hook-like structure in Br-$\alpha$ and H$_2$. The bright foreground pre-main sequence star, [MGM2012] 2301, is masked to enhance the contrast.}
    \label{fig:geometry}
\end{figure*}

\begin{figure*}[!ht]
    \centering
    \includegraphics[width=\linewidth]{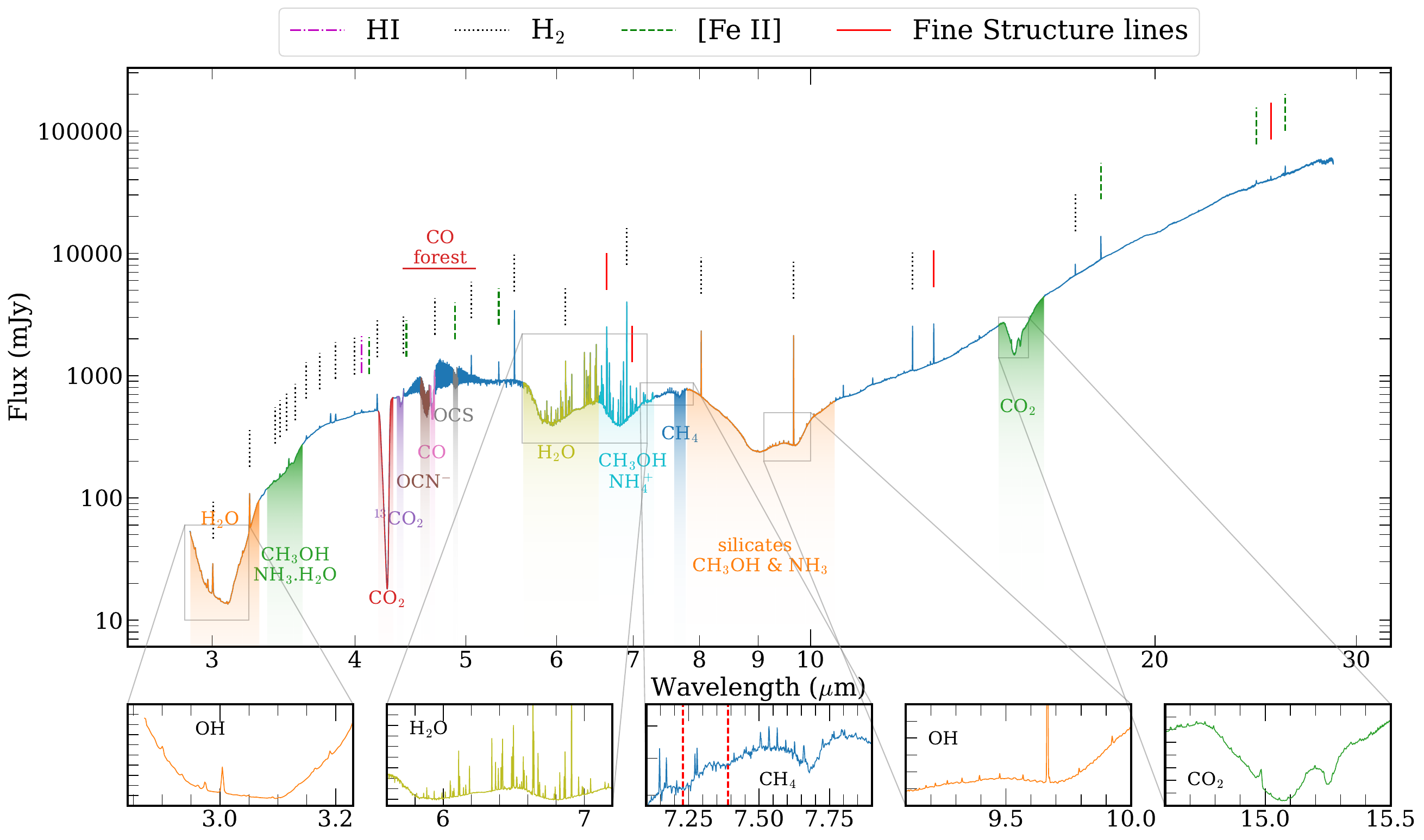}
    \caption{The extracted 1D spectrum using a 2\arcsec~aperture centered at the ALMA continuum position of HOPS 370. Prominent absorption features are gradient-shaded in colors while the detected emission lines are marked using vertical line markers. Molecular lines of \water, CH$_4$, OH, and \coo~are displayed in the insets.}
    \label{fig:1D_spec}
\end{figure*}

\section{Observations and Data reduction}
\label{obs}

The NIRSpec IFU observations of HOPS 370 were carried out on October 16, 2022,  using the F290LP/G395M filter/grating (R$\sim$700 to $\sim$1300). The spectral coverage for these observations is $\sim2.87-5.01$ ~\micron. Employing a $2\times2$ mosaicing mode with a 10\% overlap and a 4-point-dither pattern, we achieved a $6\arcsec \times6\arcsec$ field of view (FOV). Given HOPS 370's brightness, we employed the \textsc{NRSIRS2RAPID} readout pattern, incorporating 15 groups per integration and 1 integration. The effective exposure time for these observations amounted to 3501.3 s.

The raw data obtained from the Mikulski Archive for Space Telescopes (MAST) was reduced using JWST pipeline version 1.9.5.dev7+gbf7d3c9b and CRDS context files `jwst\_1069.pmap'. To achieve science-ready products we applied custom bad pixel masks and astrometry corrections. For further details regarding the data reduction and astrometry refinement processes, the reader is referred to \cite{Federman2023arXiv}.

HOPS 370 was observed on March 15, 2023, with MIRI/MRS, in all four channels covering a spectral range from $\sim4.9$~\micron~to 28~\micron\ at spectral resolving power ranging from R$\sim$4000 to 1500 \citep{Jones2023MNRAS.523.2519J}. Mosaicing settings for these observations were similar to those of NIRSpec with the 4-point-dither pattern. In MIRI observations, we have incorporated 50 groups per integration and 2 integrations with the FASTR1 readout pattern. The effective exposure time for these observations was 4484.5 s.
Dedicated observation of background with similar exposure parameters was also taken for background subtraction. The data reduction process for MIRI/MRS is discussed in \citealt{Neufeld2024arXiv240407299N}.

Our FOVs (NIRSpec and MIRI) of HOPS 370 also include another source, [MGM2012] 2301 \citep{Megeath2012AJ....144..192M}, towards the south of HOPS 370.

%% file: S3_Results.tex
\section{Results}
\label{results}

The spectral images obtained using NIRSpec and MIRI reveal the gas and dust structure of the inner 1200 au surrounding the protostar and its chemical richness
% hourglass-like protostellar geometry and chemical richness of HOPS 370
\citep{Federman2023arXiv, Adam2023arXiv, Neufeld2024arXiv240407299N}. We present the chemical inventory and spatial ice distribution in this section.

\subsection{Morphology}

Figure \ref{fig:geometry} shows the continuum and line maps of HOPS 370 that reveal the morphology of the inner protostellar environment. The star symbol in all the panels of Figure \ref{fig:geometry} marks the ALMA continuum disk position (05$\mathrm{^h}$35$\mathrm{^m}$27.63$\mathrm{^s}$ $-05^{\circ}$09\arcmin34.42\arcsec\ (ICRS); \citealt{Tobin2020ApJ...905..162T}). The highly inclined orientation (approximately $72^\circ$ inclination; \citealt{Tobin2020_LbolApJ...890..130T, Federman2023arXiv}) reveals the presence of an hourglass-shaped scattered-light cavities with the northern cavity brighter relative to the southern cavity (Figure \ref{fig:geometry}, left panel). The northern cavity shows an east-west asymmetry with brighter emission to the west.  Bright structured emission in H$_2$ is surrounded by a weaker hourglass-shaped H$_2$ emission (middle-left panel). In the northern cavity, a shocked knot is detected in both OH and H$_2$ \citep[see also][]{Federman2023arXiv, Neufeld2024arXiv240407299N}. The shocked knot is $\sim0.5$\arcsec\ north of the scattered continuum peak (see Figure \ref{fig:geometry}).
The Br-Alpha line peaks at the shock knot and the continuum source (middle-right panel). There is a fish hook-like structure in Br-$\alpha$ and H$_2$ emission in the northern cavity ($\sim2$\arcsec~away from the ALMA continuum position). 
% We have also detected a highly collimated bipolar [Fe II] jet at 5.34~\micron~
A highly collimated jet is traced in [FeII] at 5.34 \micron, this also peaks at the position of the shock knot
\citep[right panel; see also][]{Federman2023arXiv, Neufeld2024arXiv240407299N}.
This collimated jet is also seen in Br-alpha and, in the northern cavity, H$_2$ \citep{Federman2023arXiv}.

\subsection{The Chemical Inventory}

% \textcolor{red}{This section needs re-writing after Vinod's inputs.}

In Figure \ref{fig:1D_spec}, we present the combined NIRSpec and MIRI spectrum extracted from a 2\arcsec~radius aperture centered on the ALMA continuum peak (see Figure~\ref{fig:geometry}). This spectrum covers the wavelength range from 2.87~\micron\ to 28~\micron\ and reveals a rich chemical inventory in both the gas and ice phases. 

% \noindent
% \centering
% \centerline{\textbf{Gas Species}}\\
\subsubsection{Gas Species}

HOPS 370 shows a diverse set of gas-phase molecular emission lines. We detect many ro-vibrational lines from the P and R branches of CO that exhibit both extended and jet-like structures \citep{Adam2023arXiv, Federman2023arXiv}. Pure rotational lines of H$_2$ are marked in dotted black vertical lines in Figure \ref{fig:1D_spec}. We have detected pure rotational H$_2$ lines from S(1) to S(19). The H$_2$ lines appear to have originated both in the jet and its warm gas in the cavity, with emission peaking at the shock knot (Figure \ref{fig:geometry}; see also \citealt{Neufeld2024arXiv240407299N, Federman2023arXiv}; Tyagi et al. in prep.). We have also detected H$_2$O, OH, CH$_4$,  and CO$_2$ line forests, zoomed in the inset in Figure \ref{fig:1D_spec} (details about these lines will be discussed in  Manoj et al. in prep., Neufeld et al. in prep., and Rubinstein et al. in prep., see also \citealt{Neufeld2024arXiv240407299N, Francis2024A&A...683A.249F}). The HCN ro-vibrational lines from P, Q, and R branches between the wavelength range of 13.7 to 14.3~\micron\ are also detected. Further, we detect HI, [Fe II], and other fine-structure emission lines 
(see Figure \ref{fig:1D_spec}). Detailed line lists of detected molecular and atomic/ionic species are given in Table \ref{LineListMol} \& \ref{LineListAtomic}, respectively. In Table \ref{LineListMol}, we do not present lines of CO, nor the lines at NIRSpec wavelengths as they have been already presented by \citealt{Adam2023arXiv}, and \citealt{Federman2023arXiv}.

% \noindent
% \centerline{\textbf{Ice Species}}\\
\subsubsection{Ice Species}
Many ice absorption features, highlighted in Figure \ref{fig:1D_spec}, are detected at high signal-to-noise ratios (SNR). Dominant ice absorption features in the HOPS 370 spectrum include the OH stretching mode of \water\ at 3.0~\micron, CO$_2$ stretching mode at 4.27~\micron, $^{13}$CO$_2$ stretching mode at 4.38~\micron~\citep{Brunken2024}, \ocn~at 4.62~\micron, CO stretching mode at 4.67~\micron, C-O stretch mode of OCS at 4.90~\micron, H$_2$O bending mode at 6.0~\micron, CH$_3$OH/NH$_4^+$ at 6.85~\micron, CH$_4$ at 7.67~\micron, and double-peaked CO$_2$ bending mode at 15.2~\micron. We detected the 9.7~\micron~silicate absorption feature blended with the NH$_3$ umbrella and CH$_3$OH absorption features. We also detected two weak absorption features at 7.23 and 7.38~\micron~probably due to C-H deformation bands of C$_2$H$_5$OH/HCOOH, and CH$_3$CHO/HCOO$^-$ respectively (marked by red dashed lines in the third inset of the bottom row in Figure \ref{fig:1D_spec}; also see \citealt{2015Boogert, Rocha2024A&A...683A.124R} for detailed discussion on these species). Absorption bands of NH$_3$:\water, CH$_3$OH at $\sim3.5$~\micron, and HDO ice at $\sim$4.1~\micron~have also been detected \citep{Slavicinska2024arXiv240415399S}.  Additionally, we have a tentative detection of CH$_3$CN and C$_2$H$_5$CN \citep{Nazari2024arXiv240107901N}.

The presence of a narrow absorption feature at 3.1~\micron~due to crystalline H$_2$O and a double-peaked 15.2~\micron~CO$_2$ absorption feature suggests hints of thermal processing \citep{2015Boogert}. Furthermore, a double-peaked $^{13}$CO$_2$ absorption feature, attributed to ice segregation due to thermal processing, was observed by \cite{Brunken2024}. However, the origin of thermal processing remains unclear based on the spectrum at a single position.

\subsection{Spatial Variation of Ice Spectra}
\label{Spatial_Spec}

\begin{figure*}
    \centering
    \includegraphics[width=\linewidth]{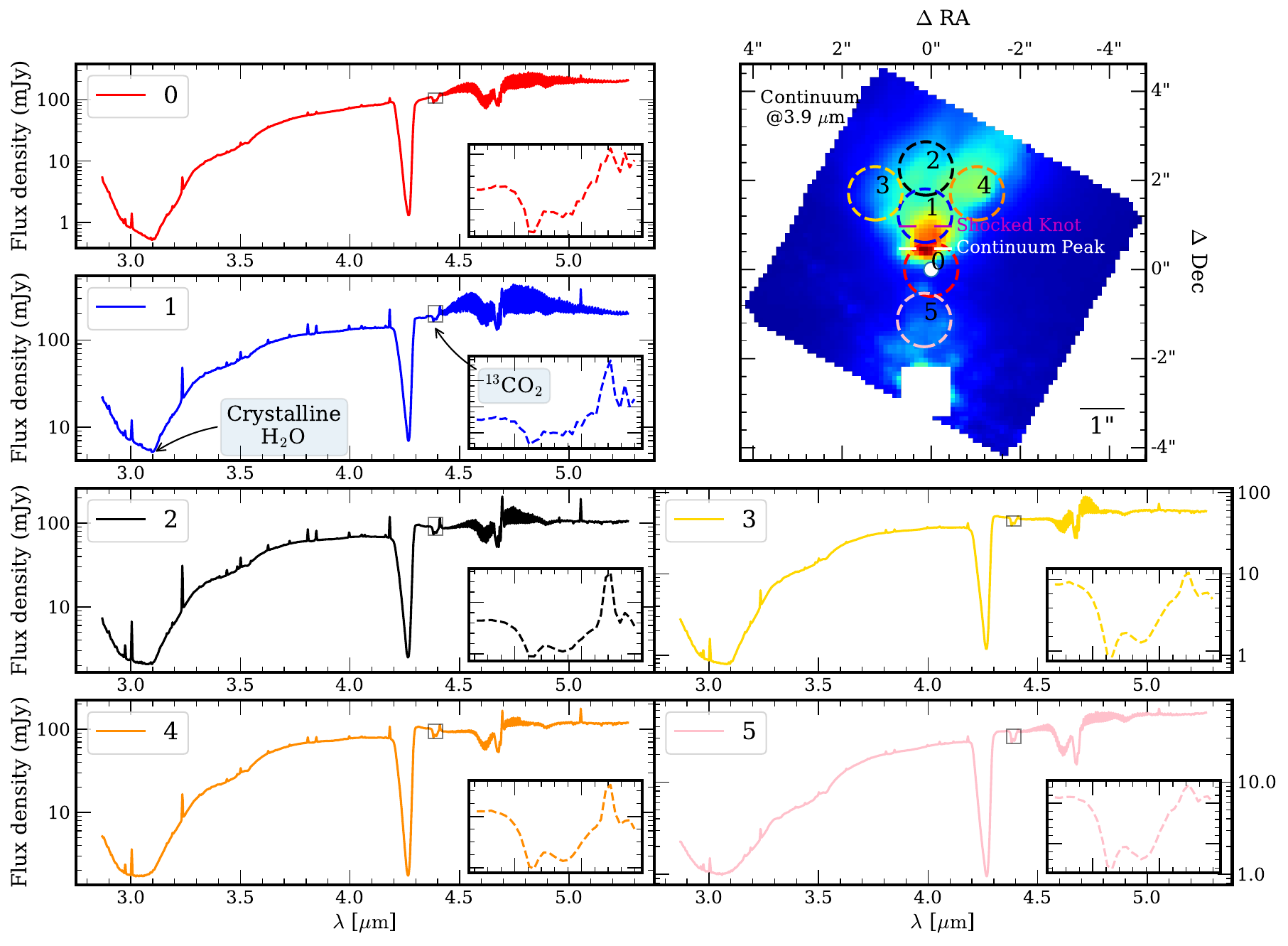}
    
    \caption{Spatial variation of the NIRSpec spectrum extracted from the circular apertures of 0.6\arcsec~radius that are shown overlaid on the 3.9~\micron~continuum image. The numbers and colors corresponding to apertures are used to point to the extracted spectrum. Other color schemes are similar to Figure \ref{fig:geometry}. Insets in the spectra panels display the \isocoo~ice feature. The [MGM2012]
    2301 is masked in the continuum image to enhance the contrast.}
    \label{fig:spatial-spec}
\end{figure*}

 The 1D NIRSpec+MIRI spectrum, centered on the protostellar position, in Figure \ref{fig:1D_spec} shows signs of thermal processing, prompting further investigation into its localization and physical origin. IFU observations offer a comprehensive dataset for this investigation, particularly in the NIRSpec observation, where extended scattered light provides an extended background continuum to be absorbed by the ice species present in the protostellar envelope (see Figure \ref{fig:geometry}, first panel from left).

To investigate the spatial variations in absorption features of ice species, we extract spectra at various positions in the protostellar envelope using circular apertures with a radius of 0.6\arcsec. Our extraction apertures are displayed in Figure \ref{fig:spatial-spec} overlaid on the 3.9~\micron~continuum image (referenced by numbers and colors in the top right panel). Extracted spectra, plotted in the sub-panels, share the same color as its extraction aperture in the right top panel. 
The overall spectra look qualitatively similar in all the apertures with some minor changes in absorption features, indicating that the dominant absorption ice species are common throughout the envelope.

The heating of \coo~ice mixtures in the envelope can sublimate the most volatile ice species (distillation) and rearrange molecular bonds (segregation), which leads to the formation of pure \coo~ice \citep[double-peaked absorption in the 4.38 and 15.2~\micron~feature indicates segregation/distillation of \coo; see also,][]{Ehrenfreund1997A&A...328..649E, Boogert2000A&A...353..349B, Pontoppidan2008}. In our observations, \isocoo~(displayed in the insets in Figure \ref{fig:spatial-spec}) also shows double-peaked absorption in all the apertures, suggesting \coo~ice distillation/segregation due to thermal processing throughout the envelope. Apertures 0, 1, 2, and 3 show the absorption peak at 3.1~\micron~due to crystalline \water. Further, we detect a relatively weaker 4.67~\micron~CO absorption with respect to 4.27~\micron~\coo~absorption feature in all the apertures. All the signatures of ice absorption features suggest that the protostellar envelope has undergone thermal processing.

\subsection{Optical Depth Maps}
\label{OD-maps}
\begin{figure*}
    \centering
    \includegraphics[width=\linewidth]{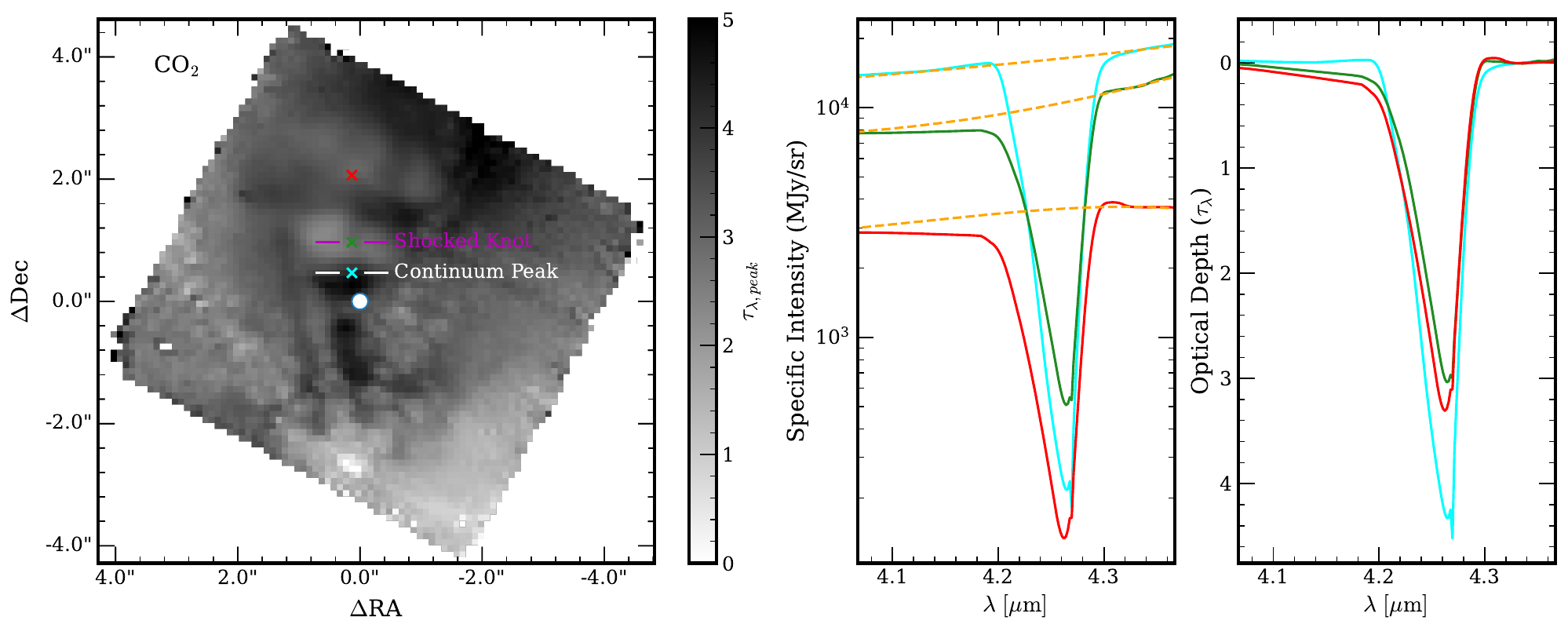}

    \caption{Peak optical depth map of \coo~and illustration of optical depth calculation for each spaxel. \textit{Left:} Peak optical depth map of \coo~at 4.27~\micron, with the white circle indicating the ALMA 870~\micron\ continuum peak. Cyan, green, and red crosses mark spaxel positions corresponding to the fit results displayed in the middle and right panels. \textit{Middle:} Specific intensity of marked spaxel (`x') as a function of wavelength. The colors of the plotted lines correspond to the same color `x' spaxel. The dashed orange lines represent the fitted continuum to the respective spaxel. \textit{Right:} Calculated optical depths for the marked spaxels.}
    \label{fig:OD427}
\end{figure*}

\begin{figure*}
    \centering
    \includegraphics[width=\linewidth]{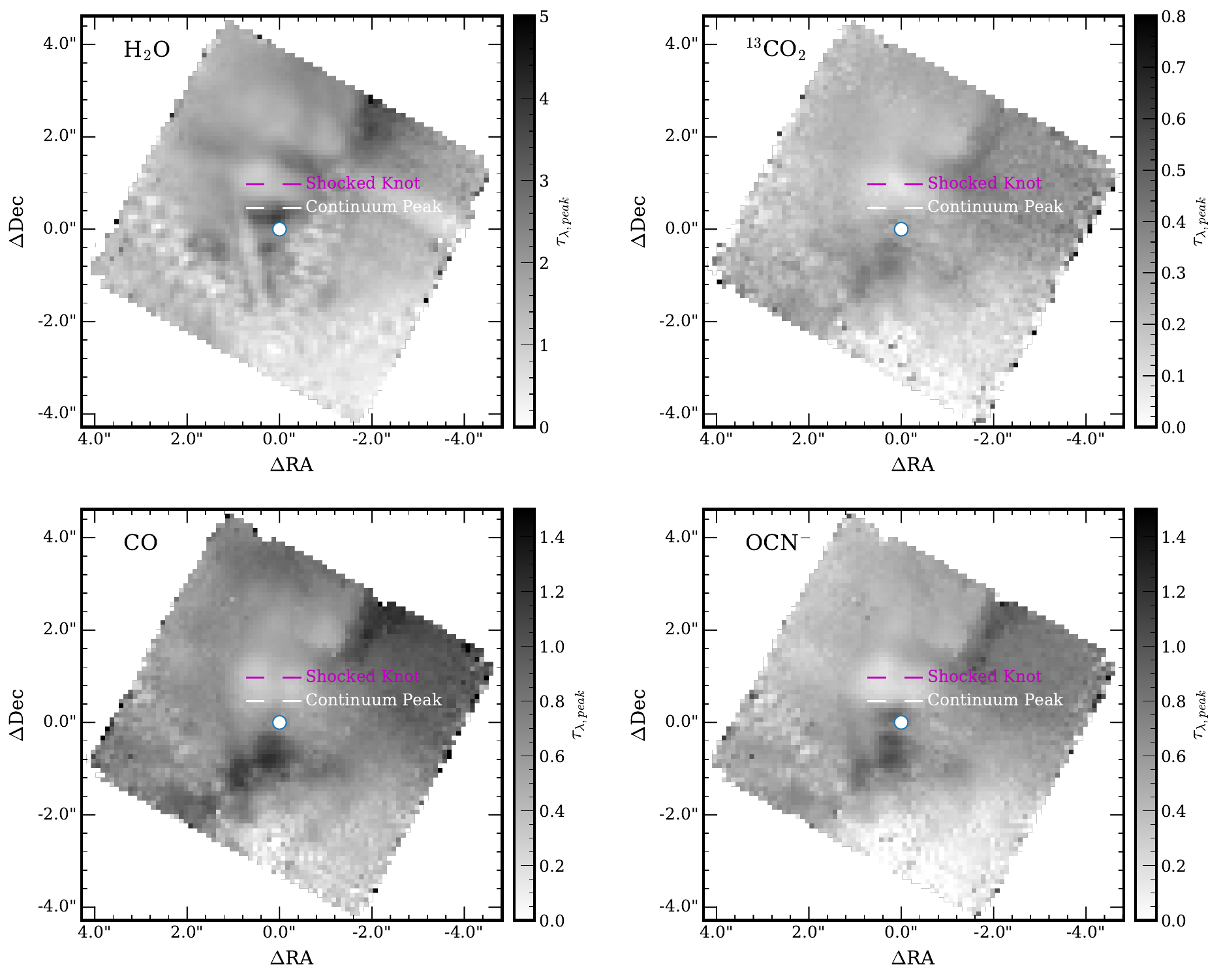}
    \caption{Peak optical depth map of \water~at 3.0~\micron, \isocoo~at 4.38~\micron, CO at 4.67~\micron, and OCN$^-$ at 4.61~\micron. The white circle marks the ALMA continuum position. The southern region of all the maps is affected by the PSF artifacts of [MGM2012] 2301.}
    \label{fig:all_ice_OD}
\end{figure*}

To investigate the spatial distribution of the ice species, we generated peak optical depth maps from the NIRSpec observations (Figures \ref{fig:OD427} and \ref{fig:all_ice_OD}).
% We first subtract all the line emission from the data cube and generate a line-free spectral cube \citep[for details, see][]{Adam2023arXiv}. The line-free spectral cube is then used to generate optical depth maps.
We first subtracted all the emission lines from the data cube and generated a line-free spectral cube \citep[for details, see][]{Adam2023arXiv}, which is then used to generate optical depth maps.

Next, we fitted a local continuum using low-order (2nd order for \isocoo, CO, and \ocn; 4th order for \coo) polynomials to the ice absorption features (except \water) at each spaxel. Local continua are used because scattering can produce large wing emissions, e.g., the blue wing of 4.27~\micron~\coo~\citep{Dartois2022A&A...666A.153D, Dartois2024NatAs...8..359D}, which makes fitting a global continuum challenging. Determining the continuum for \water~at 3.0~\micron~poses additional challenges due to incomplete spectral coverage short ward of 2.87~\micron~of our observations. To address this, a straight-line fit is applied between anchor points at 4~\micron~to 4.18~\micron~and $\sim2.87$~\micron. While fitting a straight line continuum using an anchor point at 2.87~\micron~for the \water~may underestimate the true optical depth, it facilitates comparison of relative strengths across the IFU. We calculated the optical depth at each spaxel using the following equation,
\begin{equation}
    \tau_{\lambda} = -\ln{\left[\frac{I_{\lambda,obs}}{I_{\lambda,cont}}\right]}
\end{equation}
here, $I_{\lambda, obs}$ is the observed specific intensity value and $I_{\lambda, cont}$ is the continuum estimated from the fit.

Figure \ref{fig:OD427} illustrates the results of optical depth calculations for \coo~at 4.27~\micron.
The left panel shows the peak optical depth map. The middle and right panels showcase spectral profiles with fitted continuum and the corresponding optical depth, respectively, for spaxels marked with crosses on the optical depth map. 

The peak optical depth maps of other species (\water~at 3.0~\micron, \isocoo~at 4.38~\micron, CO at 4.67~\micron, and OCN$^-$ at 4.61~\micron) are shown in Figure \ref{fig:all_ice_OD}. Peak optical depth maps of all the species show more or less similar morphology and exhibit a low optical depth region around the shocked knot (roughly 0.8\arcsec, ($\sim312$ au) north of the ALMA continuum position). There is another extended low optical depth region $\sim2$\arcsec\ ($\sim780$ au) northwest of the ALMA continuum position, separated by a band of extinction. A relatively high optical depth region is located at the scattered light continuum peak, indicating a change in scattered light emission structure at the peak ice absorption wavelength. Additionally, we find another high optical depth region, northwest of the ALMA continuum position off the edge of the outflow cavity (near $\Delta$RA, $\Delta$Dec = -2.0\arcsec, 2.0\arcsec). 

% A closer inspection of CO optical depth map reveals additional deviation of from other species near the continuum peak. 

In the southern part of the FOV, spatial information is obscured by the PSF artifacts due to [MGM2012] 2301, a bright pre-main sequence star saturated at NIRSpec wavelengths \citep[see also][]{Federman2023arXiv}. However, the \isocoo~and CO maps (top-right and bottom-left panels of Figure \ref{fig:all_ice_OD}, respectively), which are relatively less affected by artifacts, suggest an increase in optical depth, roughly following an hourglass geometry in the southern envelope similar to scattered light. 

In summary, the optical depth maps of the ice species examined show similar spatial structures in the northern cavity 
(see Figures \ref{fig:OD427} and \ref{fig:all_ice_OD}), suggesting a common origin for these structures.

\subsection{Low Extinction around Shocked Knot}
\label{lightleak}
% \begin{sidewaysfigure*}
%     \centering
%     \vspace{3.5in}
%     \includegraphics[width=\linewidth]{427_channelMaps_minimal_bar.pdf}
%     \caption{Normalized channel maps from NIRSpec observations, depicting the difference in observed intensities between continuum and \coo\ ice absorption wavelengths. The corresponding wavelength of each channel map is written on the top right. Red-shaded areas on the extracted spectra in the rightmost column of each row highlight the spectral region covered by that row's channel maps. The white `+' sign marks the ALMA continuum position. \textit{Note:} [MGM2012] 2301 is masked in this Figure.}
%     \label{fig:427ChannelMaps}
% \end{sidewaysfigure*}

\begin{figure*}
    \centering
    % \vspace{3.5in}
    \includegraphics[width=\linewidth]{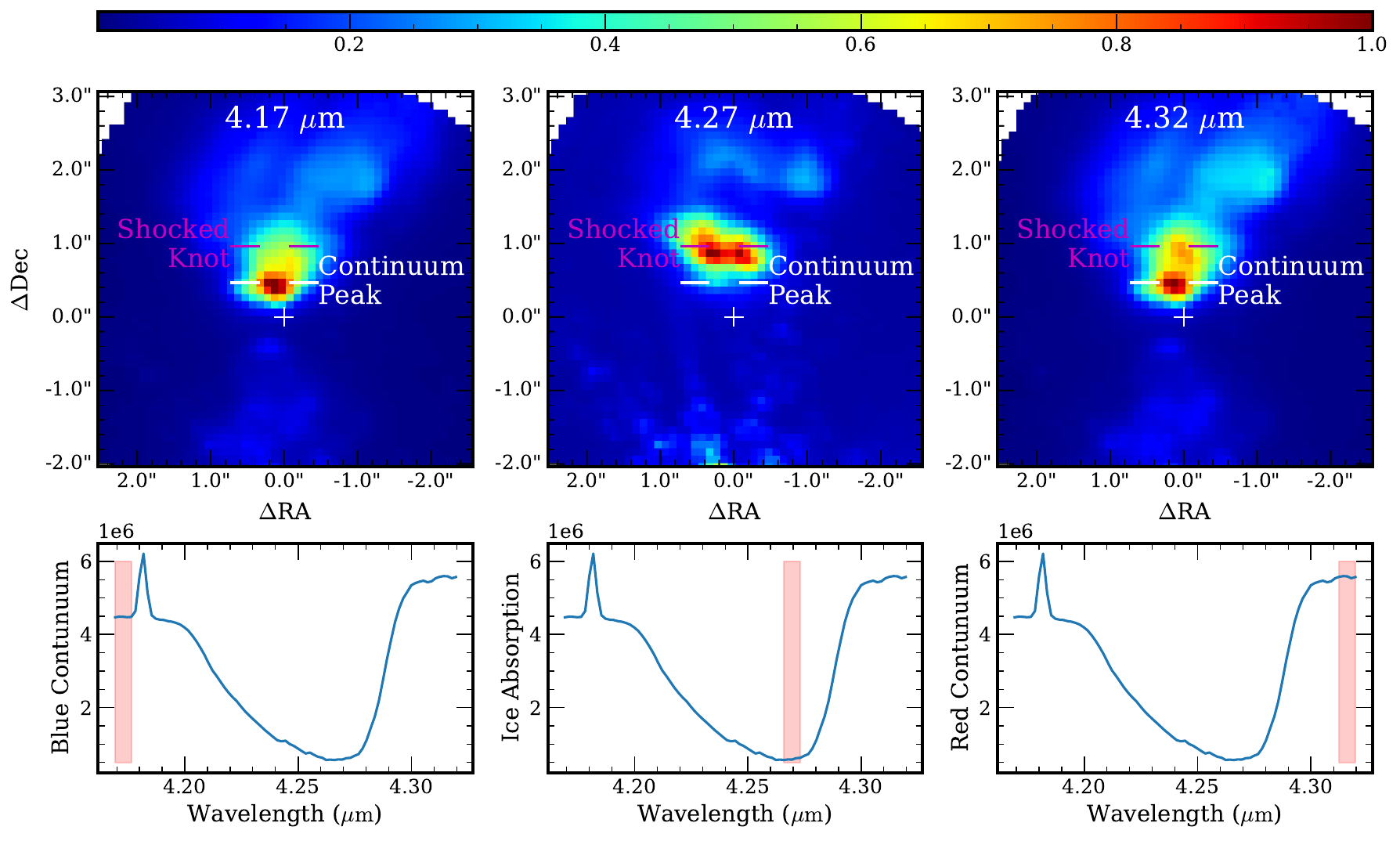}
    \caption{\textit{Top:} Normalized mean channel maps from NIRSpec observations showing the differences in observed intensities between the continuum and \coo\ ice absorption wavelengths. The mean wavelength corresponding to each map is indicated at the top of each panel. \textit{Bottom:} The red-shaded regions in the spectra below each map highlight the spectral range used to create the mean channel maps. The white ‘+’ symbol denotes the position of the ALMA continuum. \textit{Note:} The source [MGM2012] 2301 is masked in this figure.}
    \label{fig:427ChannelMaps}
\end{figure*}

Motivated by the low optical depth regions highlighted in Figures \ref{fig:OD427} and \ref{fig:all_ice_OD}, we further investigated the scattered light emission across the FOV at the peak \coo\ ice absorption wavelength (4.27~\micron), as illustrated in Figure \ref{fig:427ChannelMaps}. 
% We use channel maps to compare the configuration of the observed intensities at peak absorption wavelength with intensity configuration at continuum wavelengths on both sides of the absorption feature.
We use mean channel maps to compare the observed intensity structures at the ice absorption-free continuum and peak ice absorption wavelengths.

In Figure \ref{fig:427ChannelMaps}, the left and right column panels display the mean channel maps at ice absorption-free wavelengths on the blue and red sides of the 4.27~\micron~\coo, respectively.
The middle panel in the top row displays the observed mean channel map at the peak ice absorption wavelengths. Moreover, red-shaded regions overlaid on the extracted spectrum below each map highlight the spectral regions corresponding to the displayed mean channel maps. 
% The ALMA continuum position is marked by a white cross marker in all the channel maps.

The scattered light continuum in the left and right panels in Figure \ref{fig:427ChannelMaps} show similar hourglass geometry with the northern part being brighter. The emission peaks in both panels are marked by white horizontal bars. However, the emission structure in the middle panel differs significantly from the left and right panels, showing a shifted emission peak and disappearance of the hourglass geometry. 

A similar change in the observed intensity structures at the peak ice absorption wavelengths compared to the scattered continuum is seen for all the ice species (also evident from the optical depth maps in Figure \ref{fig:all_ice_OD}), suggesting that all the scattered light in FOV except at the shocked knot position is getting significantly absorbed at peak ice absorption wavelengths, and only the light at the shocked knot position escapes the protostellar envelope at those wavelengths.

\subsection{Mapping the Excess Crystallization of Water Ice}
\label{CrW}
\begin{figure}[ht!]
    \centering
    \includegraphics[width=\linewidth]{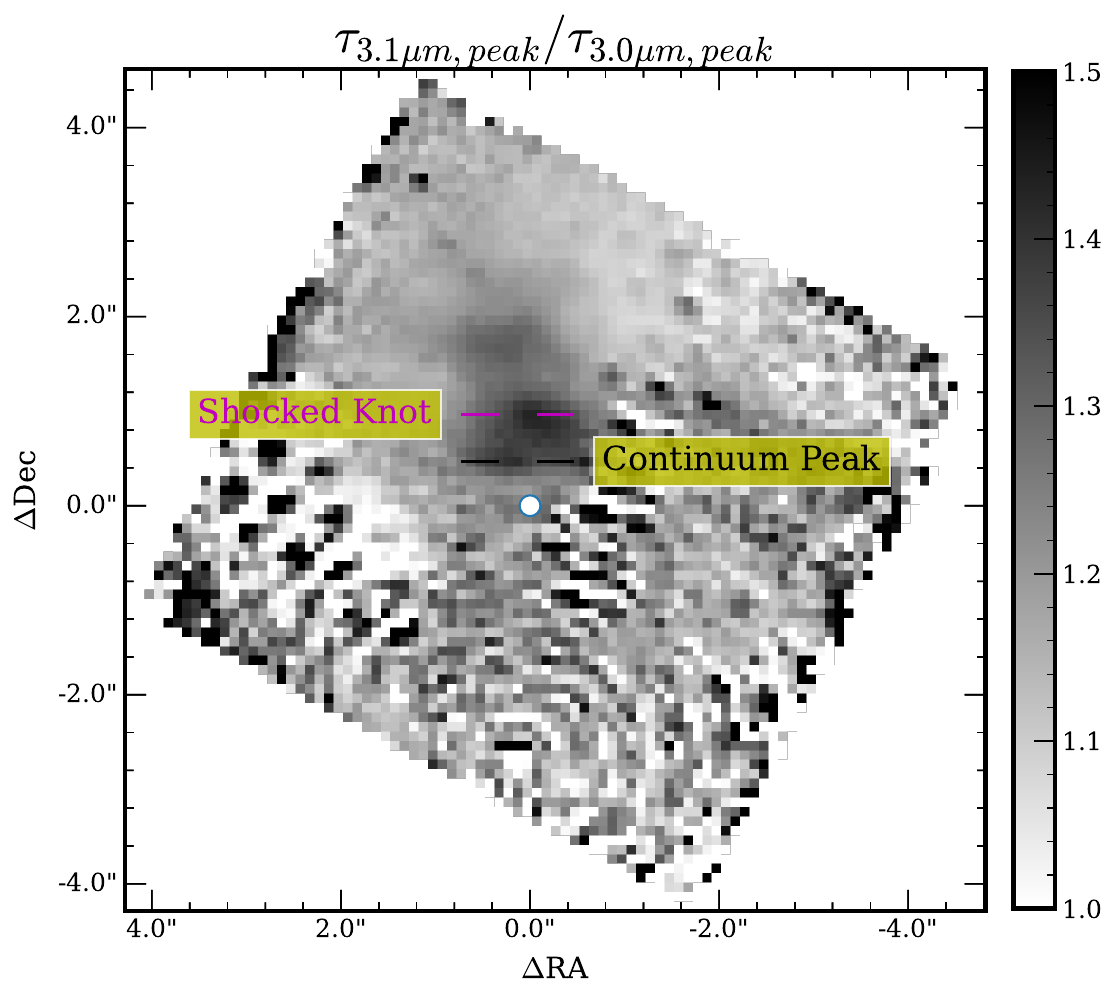}
    \caption{Ratio of peak optical depth map of \water~at 3.1~\micron~(crystalline \water) to peak optical depth map of \water~at 3.0~\micron~(amorphous \water). The white circle marks the ALMA continuum position.}
    \label{fig:Cr-water}
\end{figure}
To investigate the region of excess heating as indicated and discussed in Section \ref{Spatial_Spec}, we map the ratio between crystalline and amorphous \water~ ice within the FOV. This was done by following the procedure discussed in Section \ref{OD-maps}. Initially, optical depth maps were generated at 3.0~\micron~(the peak absorption wavelength for amorphous \water) and 3.1~\micron~(the peak absorption wavelength for crystalline \water). The ratio of the optical depth map at 3.1~\micron~to that at 3.0~\micron~then, is a measure of excess absorption due to crystalline \water. Figure \ref{fig:Cr-water} shows the ratio map, clearly indicating that the ratio is higher toward the shocked knot, with the highest water crystallization coincident with the shocked knot. The southern part of the ratio map is noisy due to the PSF artifacts of [MGM2012] 2301.

\subsection{Spatial Decomposition Maps of \texorpdfstring{\isocoo, OCN$^-$}, and CO}

\begin{figure*}
    % \centering
    % \includegraphics[width=\linewidth]{HOPS370_13CO2_decomposed.pdf}
    \gridline{\fig{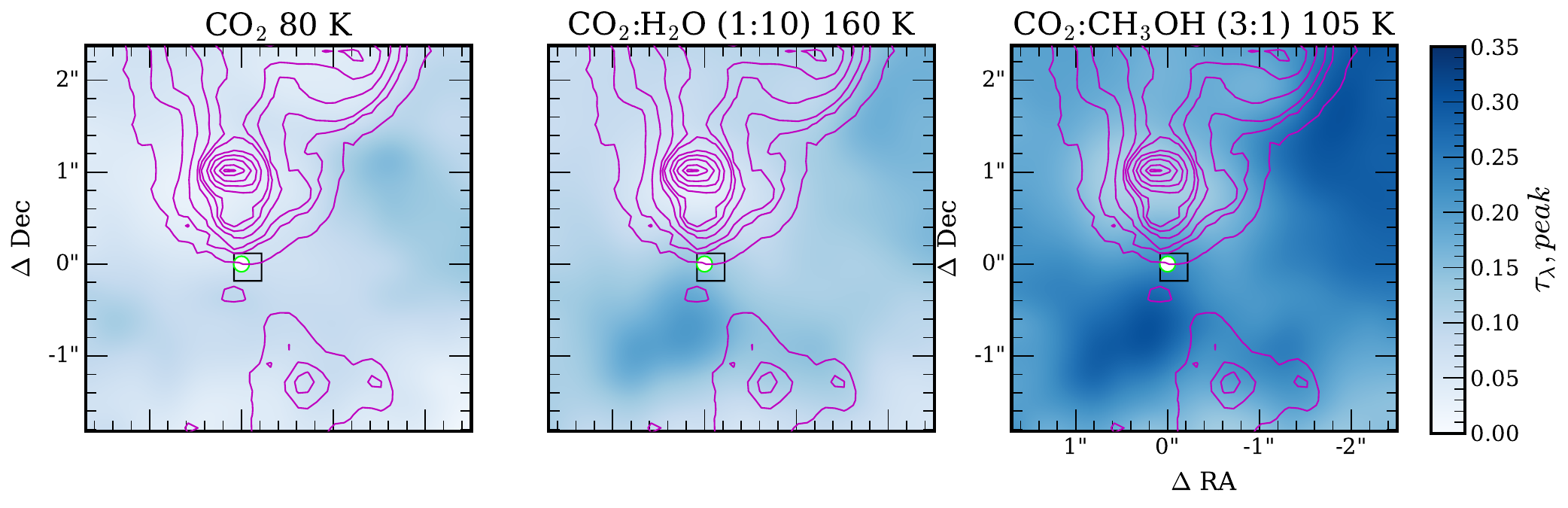}{0.75\linewidth}{[a]}
                \fig{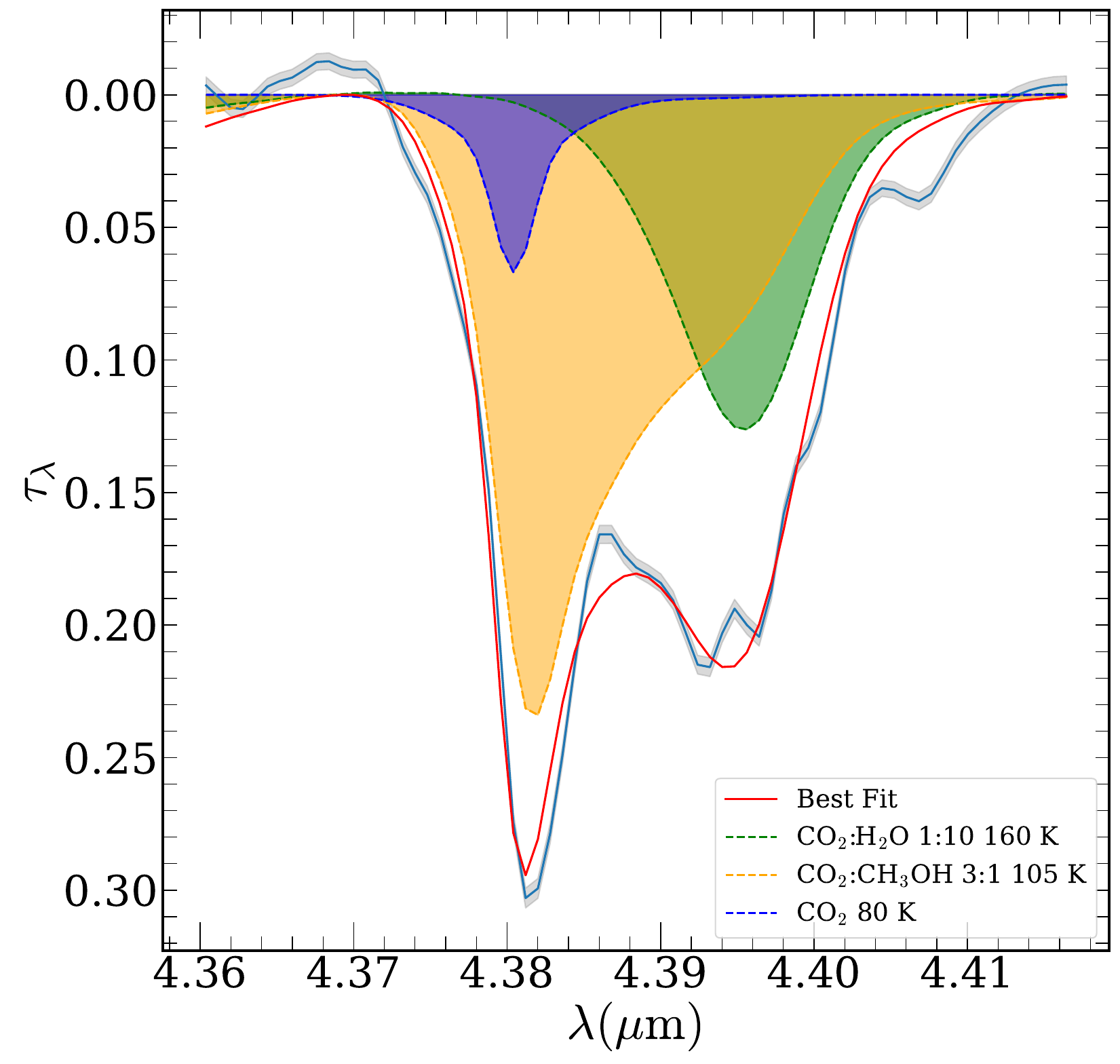}{0.25\linewidth}{[b]}}
    \gridline{\fig{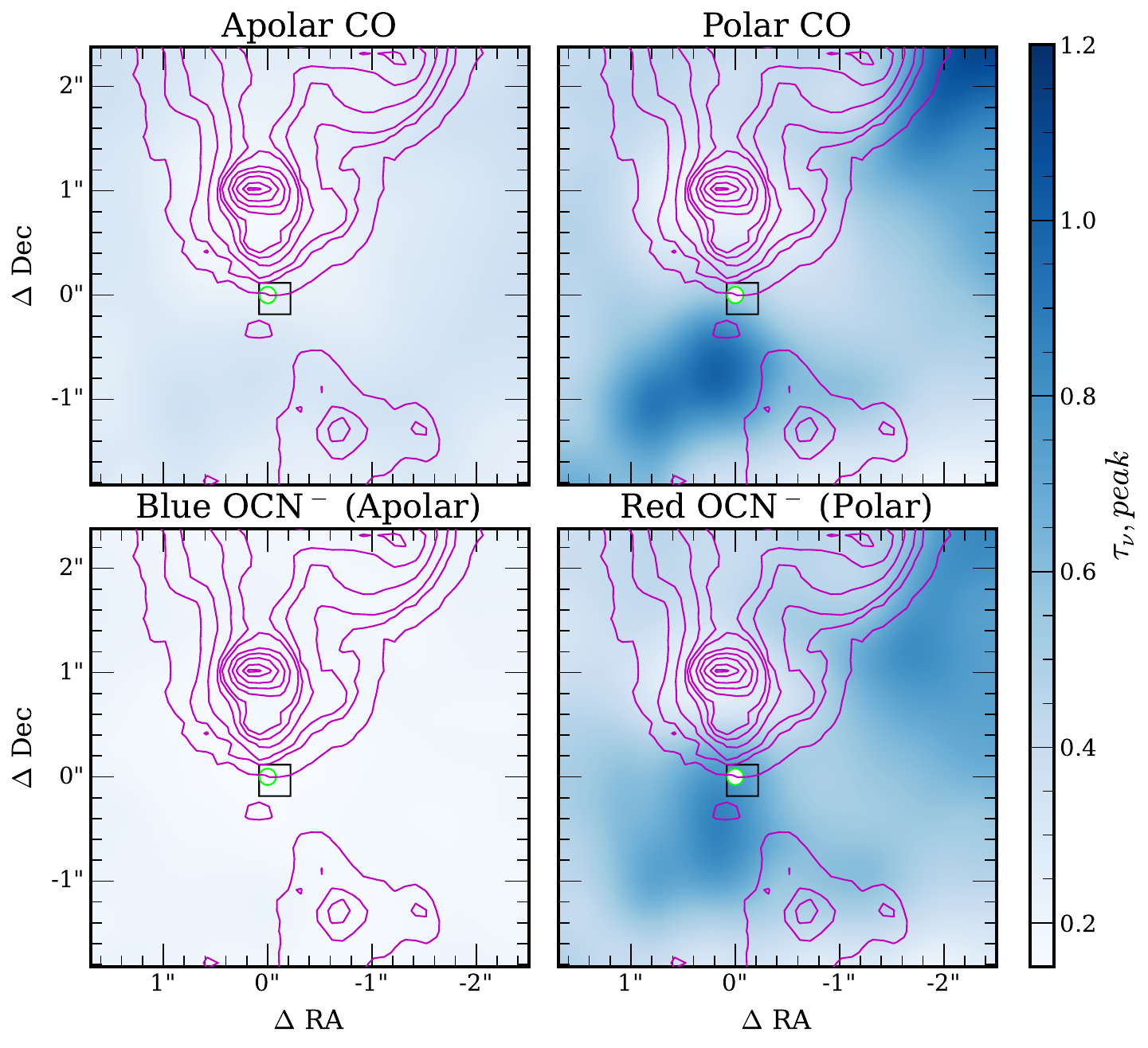}{0.5\linewidth}{[c]}
    \fig{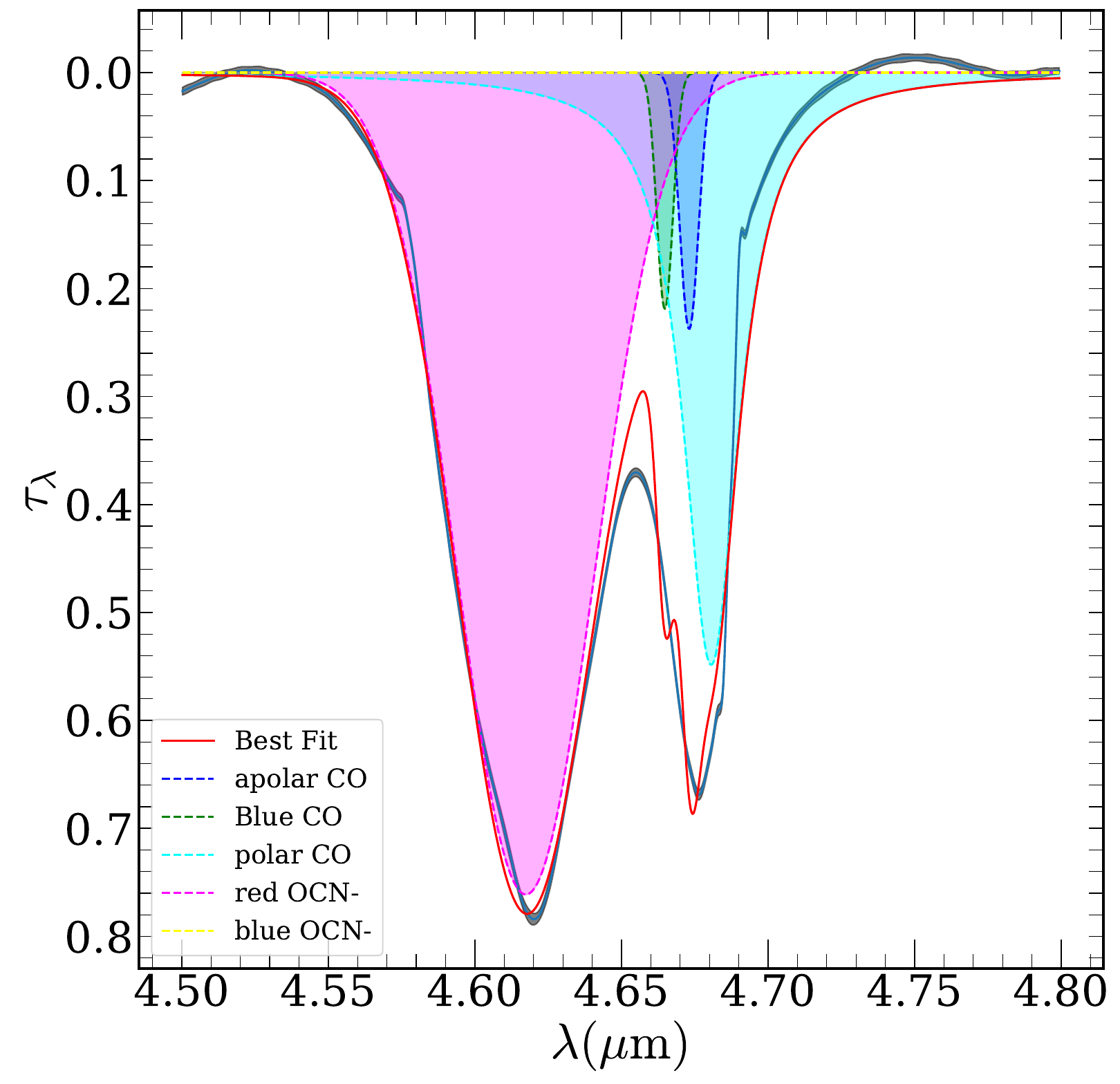}{0.4\linewidth}{[d]}}

    \caption{ \textbf{[a]:} Decomposed maps of \isocoo. The figure shows three components from left to right: \coo~at 80 K, \coo:\water~(1:10) at 160 K, and \coo:CH$_3$OH (3:1) 105 K. All panels use colormaps to represent the peak optical depth. A lime circle marks the ALMA continuum position, and the magenta color contour overlaid displays the H$_2~0-0$ S(11) emission in the FOV. The contour levels are 0.015,  0.03, 0.05, 0.1, 0.15, 0.2, 0.35, 0.5, 0.67,  0.85, 0.97$\times$(Maximum of H$_2~0-0$ S(11) emission). The densest contours (1\arcsec\ north to the ALMA continuum) mark the shocked knot. \textbf{[b]:} A fitting example of decomposition of \isocoo\ ice for the black cell marked in the peak optical depth maps in [a]. \\
    \\
    \textbf{[c]:} Decomposed maps of CO and \ocn ice. The \textit{top} row displays the apolar (left) and polar (right) components of CO, while the bottom row shows the apolar (left) and polar (right) components of \ocn. All panels use colormaps to represent the peak optical depth. The overlaid lime circle and magenta color contours are similar to those on the panel [a]. \textbf{[d]}: A fitting example of decomposition of CO and \ocn\ ice features for the black cell marked in the peak optical depth maps in [c].}
    \label{fig:13CO2Decpmposed}
\end{figure*}

% \begin{figure}
%     \centering
%     % \includegraphics[width=\linewidth]{HOPS370_13CO2_decomposed.pdf}
%     \includegraphics[width=\linewidth]{HOPS370_CO_decomposed.pdf}
%     \caption{Decomposed maps of CO and \ocn ice. The \textit{top} row displays the apolar (left) and polar (right) components of CO, while the bottom row shows the apolar (left) and polar (right) components of \ocn. All panels use colormaps to represent the peak optical depth. The lime circle marks the ALMA continuum position, and the magenta color contour overlaid depicts the H$_2~0-0$ S(11) emission in the FOV. The contour levels are 0.015,  0.03, 0.05, 0.1, 0.15, 0.2, 0.35, 0.5, 0.67,  0.85, 0.97$\times$(Maximum of H$_2~0-0$ S(11) emission).}
%     \label{fig:CODecpmposed}
% \end{figure}

To further investigate the envelope environment, we show the decomposed maps of \isocoo, OCN$^-$, and CO in Figures \ref{fig:13CO2Decpmposed}. The decomposed maps are generated by fitting the observed absorption features using laboratory spectra of contributing species in the respective ice absorption bands. These maps cover the inner 4\arcsec$\times$4\arcsec~region. 

To generate the decomposed maps, the 4\arcsec$\times$4\arcsec~region was divided into a 2D grid with a cell size of 3 pixel$\times$3 pixels ($\sim 0.3$\arcsec$\times$0.3\arcsec). This binning approach, employed instead of utilizing individual pixels, enhances the SNR for subsequent decomposition analysis.
Spectra were extracted from each 0.3\arcsec$\times$0.3\arcsec~square cell. Subsequently, we fit a low-order (3rd order for \isocoo; 2nd order for CO and \ocn) polynomial local continuum in the extracted spectra to obtain the optical depths for each cell. These derived optical depth values were then used in the decomposition of the absorption features.

For the decomposition of \isocoo, we followed the methodology outlined by \cite{Brunken2024}, where they demonstrated the necessity of including warm and hot ice species in the fitting process for HOPS 370. Specifically, we utilized laboratory spectra of pure \coo~at 80 K, \coo:\water~(1:10) at 160 K, and \coo:\methenol~(3:1) at 105 K, obtained from the Leiden Ice Database for Astrochemistry (LIDA)\footnote{\url{https://icedb.strw.leidenuniv.nl/}} \citep{Rocha2022A&A...668A..63R}, to model the observed \isocoo~absorption feature at 4.38~\micron (Figure \ref{fig:13CO2Decpmposed}[a] and \ref{fig:13CO2Decpmposed}[b]).
% Figure \ref{fig:13CO2Decpmposed}[b] shows the decomposition results for \isocoo\ ice in the black colored cell shown in Figure \ref{fig:13CO2Decpmposed}[a].

Due to the feature overlap, the decomposition of \ocn~and CO ice was done simultaneously (Figure \ref{fig:13CO2Decpmposed}[c] and \ref{fig:13CO2Decpmposed}[d]). We used a mixture of Gaussian and Lorentzian functions to fit the absorption feature \citep[see for details][and references therein]{Pontoppidan2003A&A...408..981P, vanBroekhuizen2005A&A...441..249V, Boogert2022ApJ...941...32B}. Components for both the \ocn~and CO are described in Table \ref{table_CO}. See \citealt{Nazari2024arXiv240107901N} for details about \ocn\ ice. 
% Please see Figure \ref{fig:13CO2Decpmposed}[d] for a fitting example for one cell.

\begin{deluxetable}{cccc}
\label{table_CO}
\tablecaption{CO and \ocn~Component Profiles}
\tablehead{\colhead{Component} & \colhead{Profile} & \colhead{Center (\micron)} & \colhead{FWHM (\micron)}}
\startdata
Apolar \ocn & Gaussian & 4.5969 & 0.0317 \\
Polar \ocn & Gaussian & 4.6174 & 0.0554 \\
Blue CO & Gaussian & 4.6648 & 0.0065 \\
Apolar CO & Gaussian & 4.6731 & 0.0076 \\
Polar CO & Lorentzian & 4.6806 & 0.0232 \\
\enddata
\end{deluxetable}

Figures \ref{fig:13CO2Decpmposed}[a] and \ref{fig:13CO2Decpmposed}[c] show the peak optical depth maps of various fitted components of  \isocoo, \ocn, and CO. Both figures demonstrate an excess of polar species relative to apolar species across the entire FOV. The dominance of the polar CO component relative to the apolar CO component across FOV and the requirement of hot/warm ice components to fit the \isocoo~absorption profile suggests thermal processing throughout the envelope (see Section \ref{sec:warm-discussion}). 
% The dense contours of H$_2~0-0$ S(11), located approximately 1\arcsec\ north of the ALMA continuum position and tracing the shocked knot, coincide with the low optical depth region.
The maps also show a minimum in the optical depth coincident with the bright knot in the H2 0-0 located $\sim1$\arcsec\ north of the ALMA continuum position.

%% file: S4_Discussion.tex
\section{Discussion}
\label{discussion}

The energetic processes in the protostellar phase, such as protostellar irradiation, episodic accretion-drive outbursts, and shocks driven by jets/outflows, can potentially change the chemical composition or the physical structure of the envelope material. Direct observation of these phenomena in action has been previously limited due to lack of sensitivity and angular resolution. However, the unique combination of JWST NIRSpec+MIRI's sensitivity, angular resolution, and broad spectral coverage (2.8~\micron~to 28~\micron) has finally made these detailed studies feasible. 

Our latest observations of HOPS 370 using JWST NIRSpec and MIRI reveal the chemical inventory and previously unseen details of the ice distribution within the envelope.
In this section, we discuss the results obtained in the Section \ref{results}.

% \subsection{}
% \subsection{Warm envelope: Decomposed Optical Depth Maps}
% Use section 3.3 and show decomposed maps here.

\subsection{Warm Envelope} \label{sec:warm-discussion}
As demonstrated in Section \ref{results}, multiple ice absorption signatures (e.g., weak CO ice, double-peaked 4.38~\micron~\coo, double-peaked 15.2~\micron~\coo, and the presence of crystalline water) provide evidence for significant envelope heating in HOPS 370 \citep{Pontoppidan2008, 2015Boogert, 2022Kim, Brunken2024}. Some of these signatures, e.g., double-peaked \isocoo~ice features are quite extended (see Figure \ref{fig:spatial-spec}). Additionally, CO ice absorption is much weaker relative to \coo~everywhere in the envelope (see Figures \ref{fig:OD427} and \ref{fig:all_ice_OD}). All of these together suggest thermal processing of ice species throughout the envelope in our FOV. 

Additionally, the decomposed maps of CO, and \ocn~(see Figure \ref{fig:13CO2Decpmposed}[c]) show that the polar mixtures of these ice species have higher optical depths across the envelope relative to the apolar mixtures. 
The sublimation temperatures of polar mixtures are relatively higher than those of apolar mixtures \citep[e.g., the sublimation temperatures of polar CO and apolar CO are $\sim90$ K and $\sim20$ K, respectively, see also][]{2015Boogert}. 
The combination of weak CO ice absorption, dominated by the polar mixture, along with the presence of strong ro-vibrational CO lines in emission (see Figure \ref{fig:spatial-spec}), strongly suggest the possibility of CO ice sublimation in the inner envelope up to $\sim800$ au (the extent of the decomposed maps) from the central protostar. 

\begin{figure}
    \centering
    \includegraphics[width=\linewidth]{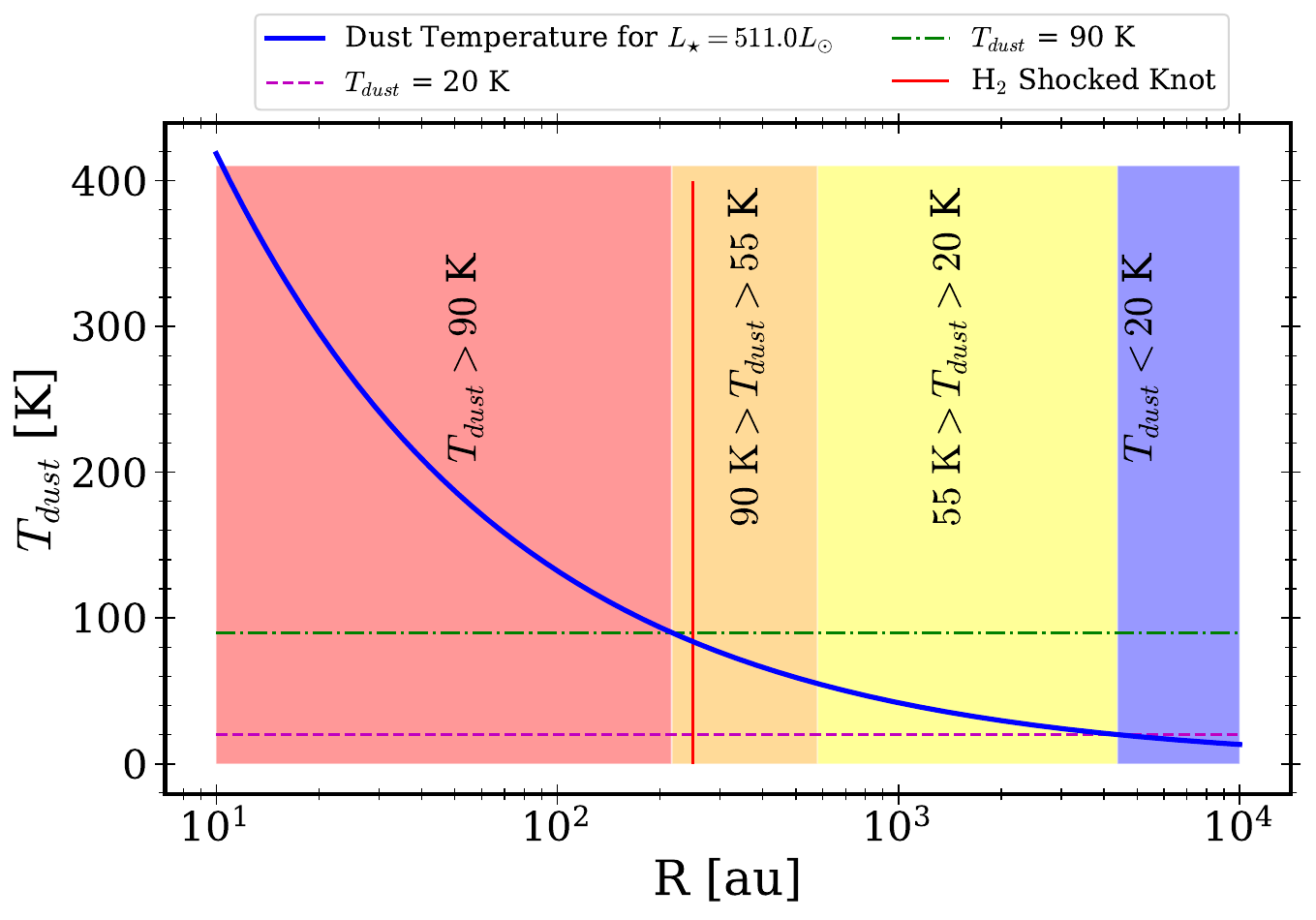}
    \caption{Radial profile of dust temperature of the HOPS 370 envelope. The color-shaded regions mark various dust temperature zones. The red vertical line represents the inclination corrected distance of the H$_2$ shocked knot from the central protostar.}
    \label{fig:RadialDustTemp}
\end{figure}

To explore the potential cause of the thermal processing of ice species over this spatial extent, we calculate the radial profile of the dust temperature in the envelope. For simplicity, we assume the envelope to be spherical and compute the equilibrium temperature of a dust grain at a distance $R$ from the central protostar. Assuming dust grains to be perfect blackbodies, the temperature of the dust grain ($T_{dust}$) at a distance $R$ from the protostar with luminosity $L_{\star}$ is given by:
\begin{equation}
    T_{dust}(R) = \left[\frac{L_{\star}}{16 \pi \sigma R^2}\right]^{1/4}
\end{equation}

Figure \ref{fig:RadialDustTemp} shows the radial profile of dust temperature. A total luminosity of 511 \lsun\ was used for this calculation \citep{Furlan2016, Tobin2020ApJ...905..162T}. The radial profile of dust temperature shows that dust may have temperatures above 20 K and 90 K for radial distances less than $\sim4000$ and $\sim200$ au, respectively. These temperatures are high enough to explain the extended ice heating observed in HOPS 370's envelope.
Hence, the envelope heating to this spatial extent in HOPS 370 can likely be attributed to its high total luminosity, $511$ \lsun, due to the large accretion rate of $\dot{M}_{acc}\sim2.25\times10^{-5}~M_{\odot}$ yr$^{-1}$ \citep{Tobin2020ApJ...905..162T}.
However, the exact mechanism responsible for the high accretion rate in HOPS 370, whether it is in a steady state or in a burst phase, is still a matter of debate \citep{Tobin2020ApJ...905..162T}.
Nevertheless, given the simplistic assumptions made in the calculations, a detailed radiative transfer modeling study is required to understand the details of the thermal structure of the envelope in the future.

\subsection{Is there a Local Ice Deficit in the Inner Envelope?}
\label{IceHoleSection}

% The rotational collapse formation of the protostar pathway suggests a smooth decrement in the envelope dust density as a function of distance from the central protostar. 

Instead of a smooth decrement in the envelope dust density as a function of distance from the central protostar, the peak optical depth maps of volatiles such as \water, \coo, \isocoo, CO, and \ocn\ reveal a highly structured spatial distribution of ice species column density in the envelope (see Figures \ref{fig:OD427} and \ref{fig:all_ice_OD}). Further, channel maps in Figure \ref{fig:427ChannelMaps} show a remarkable difference in the observed intensity distribution at peak ice absorption wavelengths compared to ice absorption-free/continuum wavelengths. The normalized channel maps show that, at peak ice absorption wavelengths, not all background scattered light is absorbed by the ices in the envelope. The scattered light leaks through an elliptical patch, spatially overlapping with the H$_2$ and OH shocked knots. These signatures suggest relatively less absorption of light by ice species at the shocked knot position. There is another low opacity bar-like region northwest of the shocked knot position ($\sim2$\arcsec~north of the ALMA continuum position).

To confirm the lack of absorption offered by ice species at these low-opacity regions, we generated extinction ($A_V$) and column density maps using the \coo~ice absorption band. To produce a column density map of \coo, we use the following equation at each spaxel,
\begin{equation}
    N_{CO_2} = \frac{1}{A} \int\tau_{\nu}d\nu
\end{equation}
where $A=1.1\times10^{-16}$ cm mol$^{-1}$ is the density-corrected band strength \citep[][]{Gerakines1995A&A...296..810G, Bouilloud2015MNRAS.451.2145B}. Next, we generate the extinction map using the KP5 extinction law \citep{Pontoppidan2024-zt}. 

\begin{figure*}
    \centering

    \includegraphics[width=\linewidth]{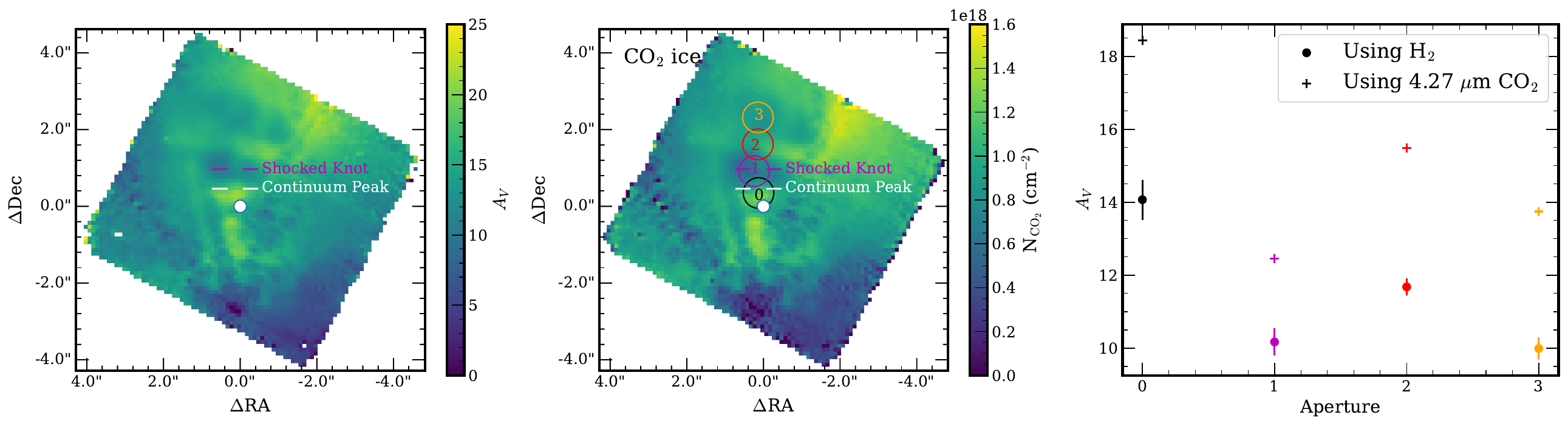}
    \caption{Extinction and column density maps using the 4.27~\micron~\coo~feature. The \textit{left} panel shows the visual extinction ($A_V$) map using KP5 and the peak optical depth of the \coo\ feature. The \textit{middle} panel shows the column density map of \coo, derived using the band strength of \coo~at 4.27~\micron. The shocked knot peak of H$_2$ and the continuum peak are marked by magenta and white color lines, respectively. The white-filled circle marks the ALMA continuum position. Black, magenta, red, and orange circles, marked by numbers 0, 1, 2, and 3 are the apertures used to calculate the extinction independently using H$_2$ rotational lines. The \textit{right} panel shows the measured extinction ($A_V$) corresponding to the apertures marked in the middle panel using H$_2$ rotational lines and 4.27~\micron~\coo~feature in cross and dot markers, respectively.}
    \label{fig:ColumnAv}
\end{figure*}

We used the relation $A_{\lambda} = 2.5 \log(e) N \kappa_{\text{ext}}(\lambda)$, where $\kappa_{\text{ext}}(\lambda)$ and $N$ represent the mass absorption coefficient and mass column density, respectively \citep{Pontoppidan2024-zt}. Then,
\begin{equation} \label{KPeq}
    \frac{A_V}{A_{\lambda}} = \frac{\kappa_{\text{ext}}(V)}{\kappa_{\text{ext}}(\lambda)}
\end{equation}
At each spaxel, we used the optical depth ($\tau_{\lambda}$) at 4.27~\micron~ of \coo~ice absorption feature to calculate $A_{\lambda}$.
Using equation \ref{KPeq}, we were able to calculate $A_V$ at each spaxel. 
% band within KP5 to compute the mass column density required to reproduce the observed peak optical depth at each spaxel. 
% We used the relation $A_{\lambda} = 2.5 \log(e) N \kappa_{\text{ext}}(\lambda)$, where $\kappa_{\text{ext}}(\lambda)$ represents the mass absorption coefficient \citep{Pontoppidan2024-zt}. By utilizing $N$, we were able to calculate $A_V$ at each spaxel. 
We present the extinction and column density maps obtained using the 4.27~\micron\ \coo\ feature in the left and middle panels of Figure \ref{fig:ColumnAv}, respectively. 
% The ALMA continuum position is marked in the white-filled circle in both maps. In both panels, the scattered light continuum peak and shocked knot are marked by the white and magenta lines. 
The shocked knot marked by magenta-colored lines coincides with the lower extinction and lower column density regions.
% Both the extinction and column density maps of \coo, displayed in Figure \ref{fig:ColumnAv}, reveal a low column density region around the shocked knot. 

We independently measure the extinction values towards the shocked knot using H$_2$ 0-0 S($J$) rotational lines (for details, see \citealt{Mayank2024, Neufeld2024arXiv240407299N}; Tyagi et al. in prep.), which provides a measure of extinction independent of ice optical depth.
Four circular apertures of radius 0.4\arcsec—labeled as 0, 1, 2, and 3 are depicted in black, magenta, red, and orange, respectively in the middle panel of Figure \ref{fig:ColumnAv} -- were used to calculate $A_V$.  We find that aperture 1, which spatially coincides with the low optical depth region and the shocked knot, exhibits a lower $A_V$ value by $\sim2$ magnitude compared to aperture 0 and 2 (see dot markers in the right panel of Figure \ref{fig:ColumnAv}). Similarly, aperture 3 has a lower extinction than aperture 0, and 2 has a higher extinction than aperture 1. The H$_2$ rotational diagrams at apertures 0, 1, 2, and 3 are presented in Appendix \ref{H2Appendix}.

In the same panel, we have marked the mean $A_V$ values of these apertures calculated using the \coo\ ice with a cross marker for comparison. Both measurements exhibit a similar trend; however, the $A_V$ values obtained from \coo\ ice are consistently higher than those obtained from the H$_2$ analysis. The difference may come from differences between the actual and adopted extinction law.  Alternatively, a weak component of foreground H2 may also lower the extinction in this band.  

Despite the difference in the absolute extinction values, this test independently verifies that indeed there is a lack of column density of material at the shocked knot position and the Br$\alpha$ fish-hook-like structure. We call this low optical depth region around the shocked knot an \textit{extinction minimum} in the envelope of HOPS 370. In Appendix \ref{IceRatioSection}, we further explore the effects of scattered light on the optical depth of \coo\ absorption feature.

Below, we explore the possible explanations for the observed substructures in the envelope ice density.

\subsubsection{Can Intense UV Radiation from the Strong Shocked Knot Produce Localized Excess Heating?}

% The complex structure of outflows as depicted by
The bright knot in the outflow, apparent in the
H$_2$ and Br$\alpha$ maps, spatially coincides with the ice column density deficiency structure of the envelope. We also observe increased relative crystallization of \water~at the shocked knot position (see Figure \ref{fig:Cr-water}) suggesting excess heating at the shocked knot position (marked by magenta lines in all the Figures). \citealt{Neufeld2024arXiv240407299N} have reported a suprathermal OH emission that peaks at the same knot position and attributed the suprathermal OH emission to the photodissociation of water by UV photons. Furthermore, the second ice deficit column density structure (roughly 2\arcsec~north of the ALMA continuum position) also spatially coincides with the fish-hook-like structure of Br$\alpha$ emission (see third panel of Figure \ref{fig:geometry}), again demonstrating the presence of UV photons at the ice deficit position.
These findings, taken together, suggest that UV photons originating from the shocked emission in jets/outflows could be responsible for excess heating of the inner envelope wall. This heating likely leads to the ice deficit through the sublimation of icy grain mantles and enhances the relative crystallization of \water\ ice. 

However in such a scenario, if we assume the shocked knot is located on the collimated jet axis, one would expect more uniform heating of the envelope instead of the heating concentrated in a localized region.
Further, the low optical depth regions show low $A_V$ values calculated using H$_2$ lines suggesting low total (ice+dust) column densities in these regions (see Figure \ref{fig:ColumnAv}). For this hypothesis to hold true, UV photons originating from the shocked knot would need also to destroy/sublimate dust. However, given the high sublimation temperature of silicate dust grains ($\sim1500$ K) and the fact that only a small fraction of UV photons escape the shocked knot region \citep{Neufeld2024arXiv240407299N}, this scenario is less likely to explain our observations.

\subsubsection{Variable Extinction due to Foreground Material or Initial Conditions}

The localized ice deficit and variable extinction observed in HOPS 370 (see Figures \ref{fig:OD427} and \ref{fig:all_ice_OD}) could be attributed to the initial conditions and collapse mechanism of its envelope, or possibly to variable foreground extinction. 

A simple case of variable foreground extinction of the molecular cloud could in principle also explain our observations. However, the protostellar envelope is the densest part along our line of sight and column density is lower by a factor of $\sim 2$ at the shocked knot (see Figure \ref{fig:ColumnAv}). Given these facts, this scenario is unlikely to explain our observations.

In the case of initial conditions, we assume that the initial core density has density irregularities and asymmetries. Then, the structures in the envelope column density might arise due to a rapid turbulent collapse of the envelope which did not give enough damping time to smooth out the initial density irregularities \citep[see also][]{Tobin2010ApJ...712.1010T}. In this scenario, the density deficit could be concentrated in specific regions, such as the inner cavity wall and the envelope's outer edge, or it could potentially exist throughout the envelope along our line of sight. Additionally, this case suggests an irregular supply of envelope material to the protostellar disk, which may affect episodic accretion due to disk instability. 
In this case, these regions would represent the initial ``seeds" that might originate the gravitational instabilities at the base of accretion bursts.

\begin{figure*}
    % \centering
    \includegraphics[width=\linewidth]{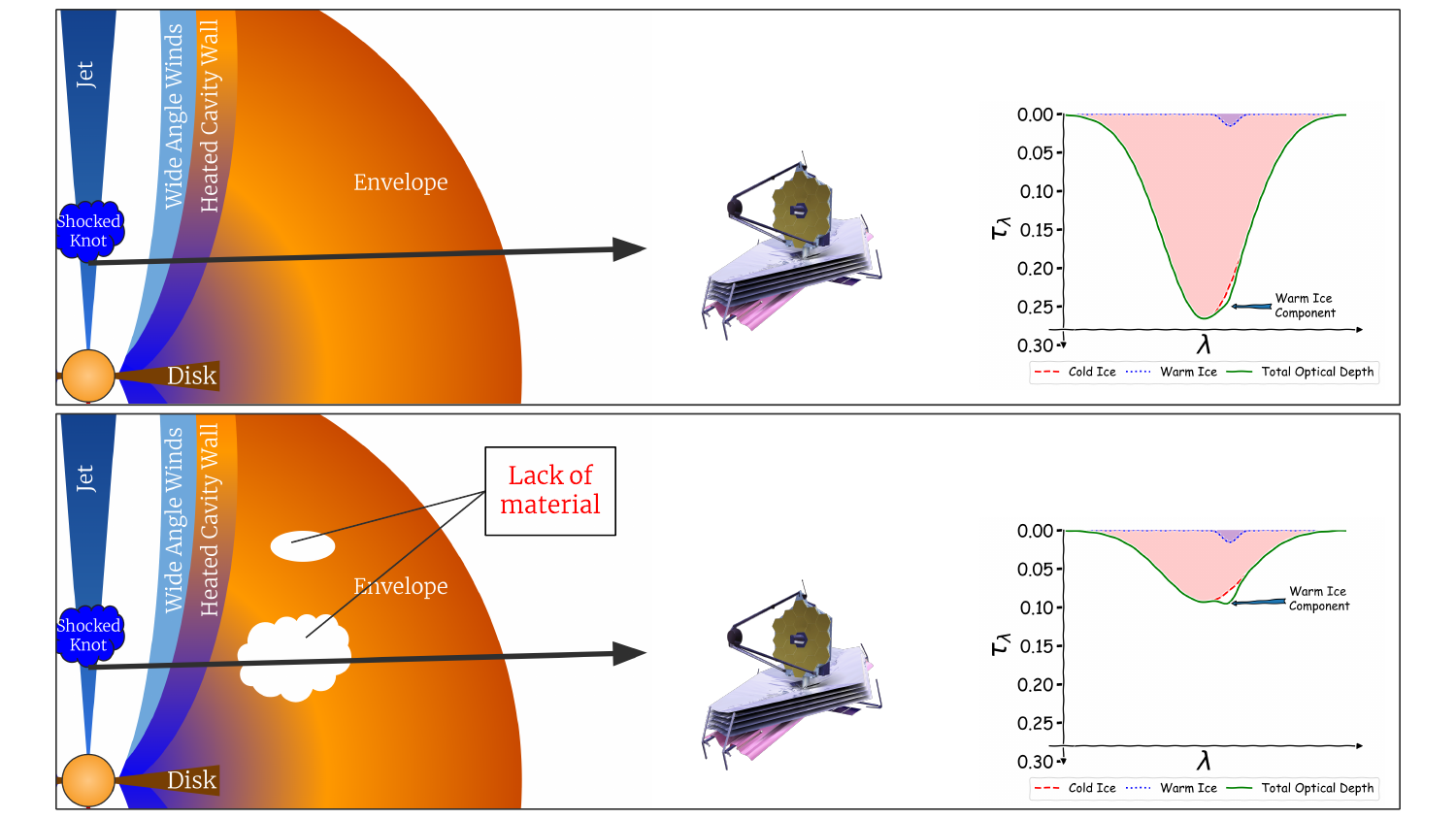}
    \caption{A simple cartoon diagram depicting the difference between theoretical (top panel) vs observed/proposed (bottom panel) density structure of the protostellar envelope. Cartoon models of the optical depth of an ice feature consisting of warm and cold ice are also shown for both cases. In the ice cartoon models, the green line shows the observed absorption profile by the JWST while the blue dotted line and green dashed lines represent the warm and cold ice components. \textit{The image is not to scale}.}
    \label{fig:Cartoon}
\end{figure*}

Assuming that the density deficit occurs in the colder part of the envelope, the relatively high crystallization of \water\ ice at the shocked knot can be explained. In this scenario, our observations would probe the inner heated regions of the envelope, while the line of sight would be depleted of the colder material in the outer parts of the envelope. This absence of colder material would manifest as localized excess heating within the envelope. Figure \ref{fig:Cartoon} displays a simplified cartoon depicting such a scenario.

%% file: S5_Concluison.tex
\section{Conclusion}
\label{conclusion}

Our JWST NIRSpec IFU and MIRI MRS observations of the intermediate-mass protostar HOPS 370 (OMC2-FIR3) demonstrate that during the protostellar phase the envelope can be significantly heated, thereby providing conditions for further processing of building material for the protostellar/protoplanetary disk. 
This study presents a rich inventory of gas-phase and ice-phase species in the spectral range from 2.8~\micron~to 28~\micron.
Further, we present the first high spatial resolution ($\sim80$ au) maps of ice species in the protostellar envelope using NIRSpec data. Below we summarize our key findings from these observations:
\begin{enumerate}
    \item The envelope of HOPS 370 exhibits clear evidence of thermally processed ices, seen in various ice absorption features: crystalline \water, double-peaked \isocoo, and the double-peaked 15.2~\micron\ \coo.
    
    \item The decomposed maps of CO and \ocn\ show that the polar mixture of ice species dominates the entire inner envelope, indicating additional evidence for thermal processing. Weak CO ice absorption compared to \coo\, together with strong CO ro-vibrational gas phase lines further suggest CO ice sublimation in the envelope.

    \item Optical depth and column density maps of volatile ice species reveal a nonuniform and structured density distribution of the protostellar envelope with a minimum in ice column density and line-of-sight extinction coincident with a prominent shock heated knot in an outflow jet.
    
    \item Dust extinction measured with H$_2$ rotational lines show a region of lower extinction, and hence lower dust column density region, at the shocked knot position.

    \item The enhanced relative crystallization of \water~observed around the shocked knot position suggests local excess heating or deficit of outer colder envelope material along the line of sight.

    \item 
    We propose that the irregular structure in the envelope has resulted in a lower column density/extinction line of sight, and the observed emission along this sightline is dominated by the inner thermally processed ices near the outflow cavity wall. This shows clear evidence of the thermal processing of the ice closer to the inner cavity wall. The spatial coincidence of the shocked knot seen in H$_2$ and Br$\alpha$ emission along this line of sight appears to be a chance alignment.

\end{enumerate}

Although the high luminosity of HOPS 370 dominates the envelope heating, the role and contribution of the outflow shocks in thermal processing of the protostellar envelope have not been fully explored yet. Future observations of low-luminosity protostars (where protostellar luminosity is inefficient in heating the envelope) with JWST may reveal the role of outflow shocks in the thermal processing of ices in the envelope.

%% file: S6_7_Acknowledgment.tex
\section{Data availability}
{All of the data presented in this article were obtained from the Mikulski Archive for Space Telescopes (MAST) at the Space Telescope Science Institute. The specific observations analyzed can be accessed via \dataset[DOI: 10.17909/3kky-t040]{https://doi.org/10.17909/3kky-t040}.}

\section{Acknowledgment}
This work is based on observations made with the NASA/ESA/CSA James Webb Space Telescope. The data were obtained from the Mikulski Archive for Space Telescopes at the Space Telescope Science Institute, which is operated by the Association of Universities for Research in Astronomy, Inc., under NASA contract NAS 5-03127 for JWST. These observations are associated with program \#1802. H.T. and P.M. acknowledge the support of the Department of Atomic Energy, Government of India, under Project Identification No. RTI 4002. Support for SF, AER, STM, RG, WF, JG, JJT, and DW in program \#1802 was provided by NASA through a grant from the Space Telescope Science Institute, which is operated by the Association of Universities for Research in Astronomy, Inc., under NASA contract NAS 5-03127. D.A.N. was supported by grant SOF08-0038 from USRA. A.C.G. has been supported by PRIN-MUR 2022 20228JPA3A “The path to star and planet formation in the JWST era (PATH)” and by INAF-GoG 2022 “NIR-dark Accretion Outbursts in Massive Young stellar objects (NAOMY)”. G.A. and M.O., acknowledge financial support from grants PID2020-114461GB-I00 and CEX2021-001131-S, funded by MCIN/AEI/10.13039/501100011033. Y.-L.Y. acknowledges support from Grant-in-Aid from the Ministry of Education, Culture, Sports, Science, and Technology of Japan (20H05845, 20H05844, 22K20389), and a pioneering project in RIKEN (Evolution of Matter in the Universe). {Leiden astrochemistry thanks support from the European Research Council (ERC) under the European Union’s Horizon 2020 research and innovation programme (grant agreement No. 101019751 MOLDISK). AS gratefully acknowledges support by the Fondecyt Regular (project
code 1220610), and ANID BASAL project FB210003.}

%% file: S8_Appendix.tex
\appendix
\restartappendixnumbering

\section{Line List}
\label{LineListTable}
\input{Line_List_Mol}
\input{Line_List_ion}

\section{H\texorpdfstring{$_2$}~ Rotational diagrams}
\label{H2Appendix}
\begin{figure}[ht!]
    \centering
    \gridline{\fig{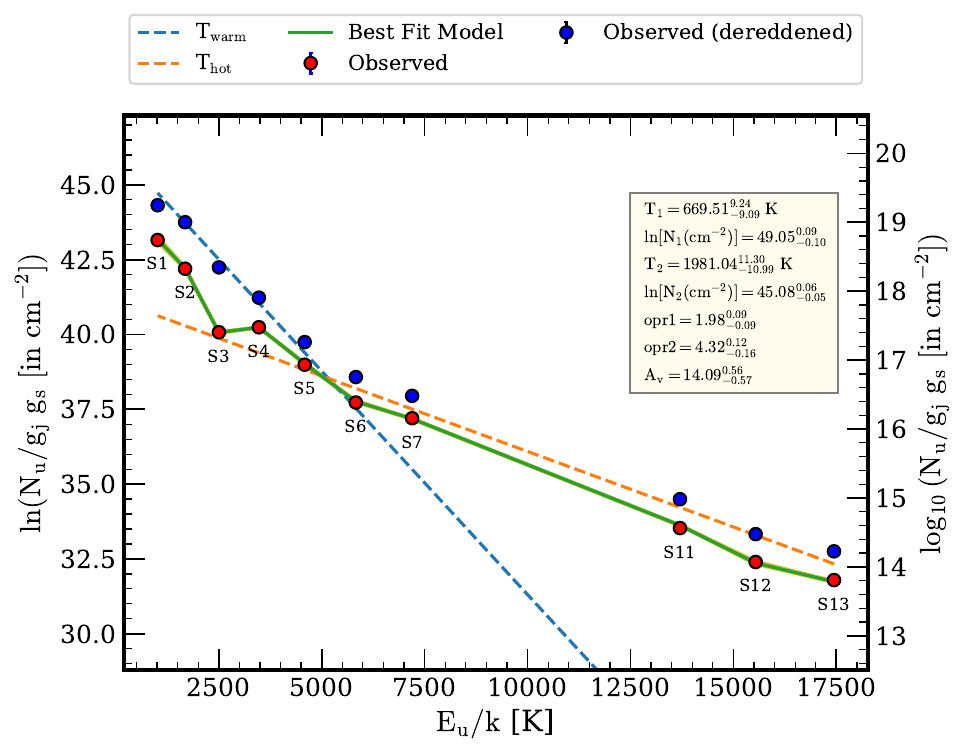}{0.5\textwidth}{(0)}
    \fig{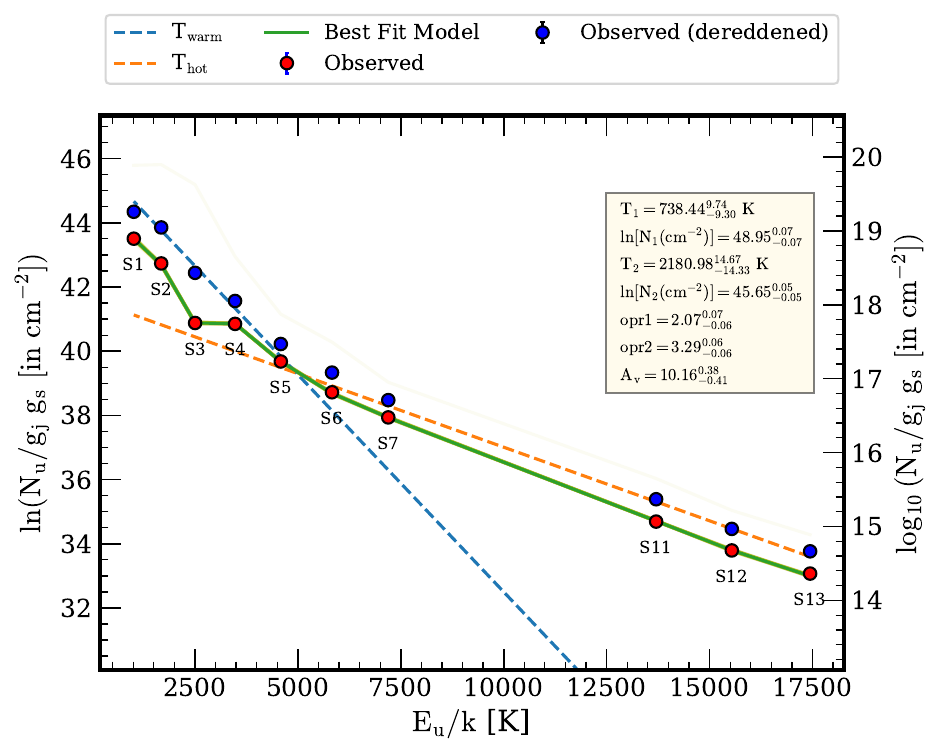}{0.5\textwidth}{(1)}}
    \gridline{\fig{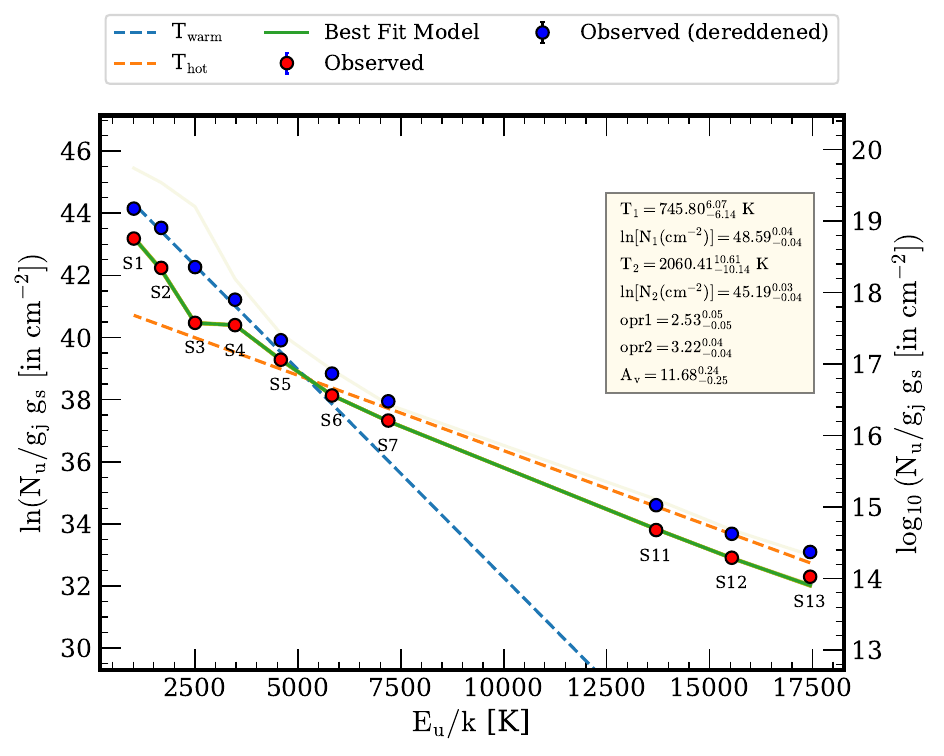}{0.5\textwidth}{(2)}
    \fig{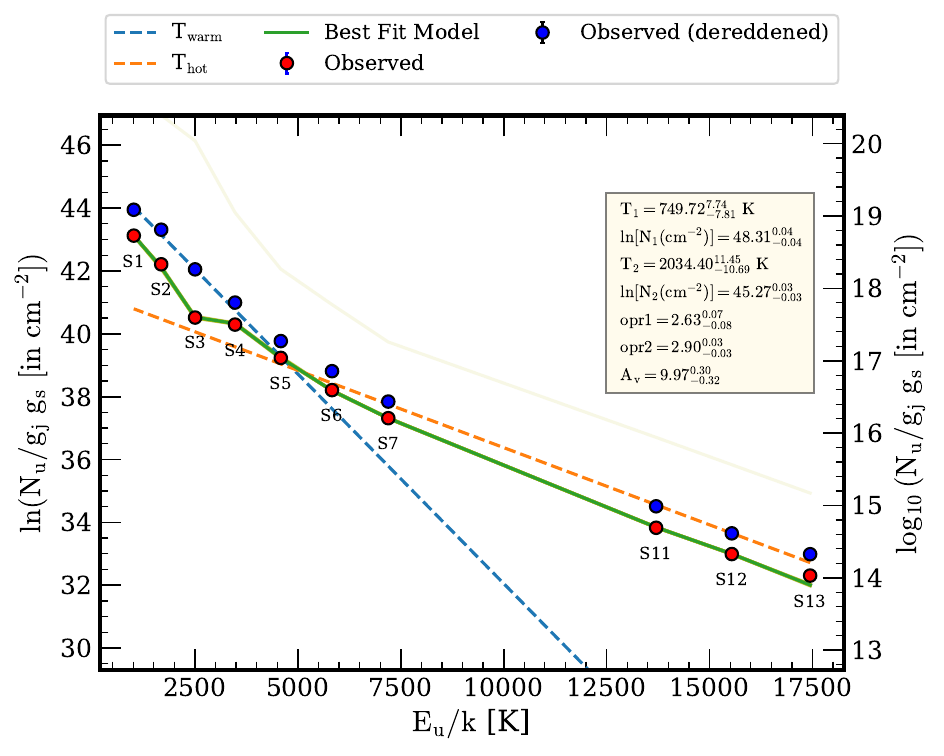}{0.5\textwidth}{(3)}}    
    \caption{H$_2$ rotational diagram for apertures 0, 1, 2, and 3 displayed in Figure \ref{fig:ColumnAv}. Red points are the observed data points while the blue points are de-reddened points using the KP5 extinction law. Blue and orange dashed lines correspond to cold and hot temperature gas considered in the fitting model. The green line is the best fit for the observed rotational diagram.}
    \label{fig:enter-label}
\end{figure}

\section{\texorpdfstring{\coo~and \isocoo}~~Ratio Maps}
\label{IceRatioSection}
\begin{figure}
    \centering
    \includegraphics[width=0.8\linewidth]{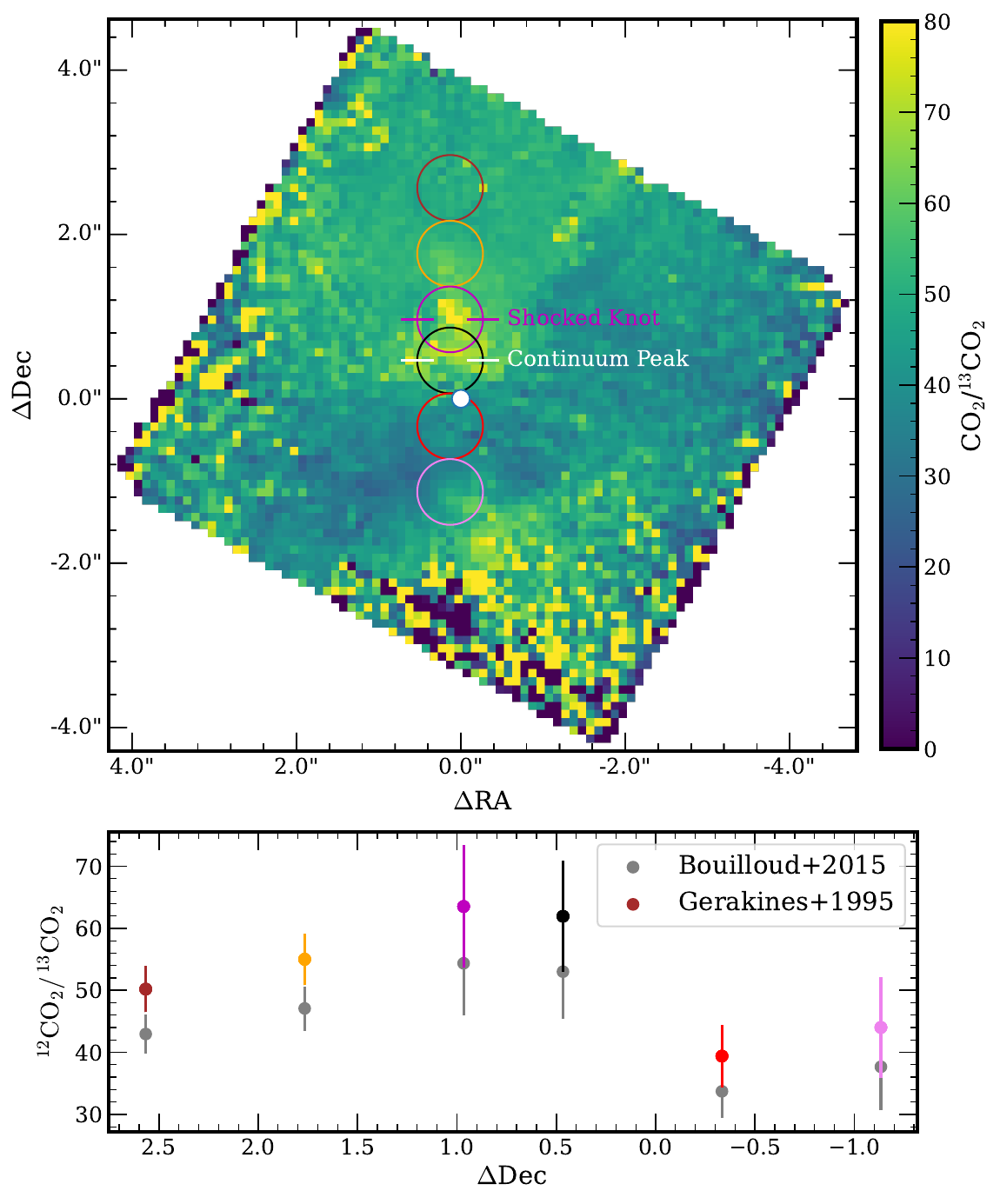}
    \caption{\textit{Top panel:} \coo~(4.27~\micron) to \isocoo~(4.38~\micron) ratio map. Shocked knot and continuum peaks are marked by magenta and white color lines, respectively. Colored apertures marked on the map are used to calculate the mean \coo/\isocoo~ratio plotted in the bottom panel. \textit{Bottom panel:} Mean \coo/\isocoo~ratio of marked apertures in the top panels as a function of the angular distance of aperture from the ALMA continuum position. Multi-colored and gray markers represent ratio values calculated using the density corrected band strength values from \citealt{Gerakines1995A&A...296..810G}, and \citealt{Bouilloud2015MNRAS.451.2145B}, respectively. }
    \label{fig:CO2_ratioMap}
\end{figure}

% To investigate the \coo/\isocoo~ratio in HOPS 370's envelope, 
To further investigate the nature of the hole in the \isocoo\ absorption, we examine the \coo/\isocoo~ratio in HOPS 370's envelope.  We generated a \isocoo~column density map using a band strength (density-corrected), $A=1.15\times10^{-16}$ cm/mol \citep[][similar to the \coo~column density map discussed in Section \ref{IceHoleSection}]{Gerakines1995A&A...296..810G, Bouilloud2015MNRAS.451.2145B}. The column density ratio map of \coo/\isocoo~is shown in the top panel of Figure \ref{fig:CO2_ratioMap}. 
We calculated the mean molecular ratios of \coo/\isocoo~in the aperture marked by magenta and black colors at the shocked knot and the continuum peak in Figure \ref{fig:CO2_ratioMap}, yielding values of $63.5\pm9.8$ and $62.0\pm8.9$, respectively.
Moving away from the shocked knot in both the north and south directions, we observed a decreasing trend in the \coo/\isocoo~ratios (see bottom panel of Figure \ref{fig:CO2_ratioMap}).

The \coo/\isocoo\ ratio can vary depending on the band strengths used in the column density calculation. Figure~\ref{fig:CO2_ratioMap} (bottom panel) compares ratios calculated with band strengths from \citealt{Gerakines1995A&A...296..810G} (density-corrected values; shown in multiple colors) and \citealt{Bouilloud2015MNRAS.451.2145B} (in gray; $A=6.8 \times 10^{-17}$ cm/mol for \coo\ and $A=7.6 \times 10^{-17}$ cm/mol for \isocoo). The mean ratio difference between the two sets is a factor of 1.17. For this discussion, we adopt the density-corrected band strengths of \citealt{Gerakines1995A&A...296..810G} as \citealt{Bouilloud2015MNRAS.451.2145B} suspected that their band strength values are lower due to possible saturation of 4.27 \coo\ absorption feature. For a detailed discussion on the band strength uncertainties, the reader is referred to Brunken et al. (in prep).

The obtained molecular ratios of \coo/\isocoo~in HOPS 370's envelope, closer to the continuum peak and shocked knot, are similar to the average local ISM ratio of 69 \citep{Boogert2000A&A...353..349B}, as well as the local dark cloud ratio of 69--87 measured in the Chameleon I star-forming region \citep{2023McClure}. However, farther from the continuum peak and shocked knot the ratio decreases by a factor of $\sim1.3-1.7$.
It is worth noting that \citealt{Boogert2000A&A...353..349B} reported variations in the \coo/\isocoo~ratio in some sources. Along the lines of sight toward GC 3, W 33A, and HH 100, they measured low \coo/\isocoo~ratios of 52, 53, and 52, respectively. 

Alternatively, the variation in the ratio may result from radiative transfer effects.\citealt{Sturm2023A&A...679A.138S} suggested that the saturation of the main isotopologue of \coo\ ice absorption band by scattered light could lead to a lower \coo/\isocoo~ratio. They observed a molecular ratio of 14 in the edge-on protoplanetary disk HH 48 NE. As mentioned in Section \ref{results}, our optical depth calculations also utilize scattered light as a background continuum. Therefore, it is plausible that low \coo/\isocoo~ratios farther away from the central protostar in our observations may be attributed to the partial saturation of the \coo~main isotopologue. This can lower the observed optical depth of the \coo\ feature, resulting in an underestimate of the column density in the protostellar envelope. 

Despite this potential issue with the \coo\ optical depth. The presence of the low opacity hole is clearly supported by other evidence.  The hole is seen in multiple ice including \isocoo\ and \water. In addition, the independent extinction measurement using H$_2$ rotational lines agrees with the low extinction at the shocked knot position.

% \section{Decomposed fitting examples}\label{Fitting_Appendix}

%% file: Line_List_Mol.tex
%% The values (usually only l,r and c) in the last part of
%% \begin{deluxetable}{} command tell LaTeX how many columns
%% there are and how to align them.

\startlongtable
\begin{deluxetable}{cccc}
\label{LineListMol}

\tablecaption{Detected Molecular lines. Molecular line information is taken from HITRAN \citep{Gordon2022JQSRT.27707949G}.}
\tablehead{\colhead{Species} & \colhead{Wavelength} & \colhead{Transition, v} & \colhead{Transition, J}\\ 
\colhead{} & \colhead{(\micron)} & \colhead{} & \colhead{}} 

%% All data must appear between the \startdata and \enddata commands
\startdata
OH & 2.9343 & X3/2 1 -- X3/2 0 & PP 4.5ee \\
OH & 2.9346 & X3/2 1 -- X3/2 0 & PP 4.5ff \\
OH & 2.9603 & X1/2 1 -- X1/2 0 & PP 4.5ff \\
OH & 2.9604 & X1/2 1 -- X1/2 0 & PP 4.5ee \\
OH & 2.9700 & X3/2 1 -- X3/2 0 & PP 5.5ee \\
OH & 2.9704 & X3/2 1 -- X3/2 0 & PP 5.5ff \\
OH & 3.0415 & X1/2 1 -- X1/2 0 & PP 6.5ff \\
OH & 3.0418 & X1/2 1 -- X1/2 0 & PP 6.5ee \\
OH & 3.0481 & X3/2 1 -- X3/2 0 & PP 7.5ee \\
OH & 3.0487 & X3/2 1 -- X3/2 0 & PP 7.5ff \\
OH & 3.0850 & X1/2 1 -- X1/2 0 & PP 7.5ff \\
OH & 3.0855 & X1/2 1 -- X1/2 0 & PP 7.5ee \\
OH & 3.0906 & X3/2 1 -- X3/2 0 & PP 8.5ee \\
OH & 3.0913 & X3/2 1 -- X3/2 0 & PP 8.5ff \\
OH & 3.1354 & X3/2 1 -- X3/2 0 & PP 9.5ee \\
OH & 3.1363 & X3/2 1 -- X3/2 0 & PP 9.5ff \\
H$_2$ & 5.0530 & 0 -- 0 & S(8) \\
H$_2$ & 5.5116 & 0 -- 0 & S(7) \\
H$_2$ & 5.8112 & 1 -- 1 & S(7) \\
H$_2$O & 5.8608 & 010 -- 000 & 2 2 0 -- 1 1 1 \\
H$_2$O & 5.8830 & 010 -- 000 & 2 2 1 -- 1 1 0 \\
H$_2$O & 5.8967 & 010 -- 000 & 5 0 5 -- 4 1 4 \\
H$_2$O & 5.9356 & 010 -- 000 & 4 1 4 -- 3 0 3 \\
H$_2$O & 5.9701 & 010 -- 000 & 4 0 4 -- 3 1 3 \\
H$_2$O & 5.9904 & 010 -- 000 & 3 1 3 -- 2 0 2 \\
H$_2$O & 6.0142 & 010 -- 000 & 2 2 1 -- 2 1 2 \\
H$_2$O & 6.0490 & 010 -- 000 & 2 1 2 -- 1 0 1 \\
H$_2$O & 6.0521 & 010 -- 000 & 3 0 3 -- 2 1 2 \\
H$_2$O & 6.0670 & 010 -- 000 & 2 2 0 -- 2 1 1 \\
H$_2$O & 6.0757 & 010 -- 000 & 3 2 1 -- 3 1 2 \\
H$_2$ & 6.1088 & 0 -- 0 & S(6) \\
H$_2$O & 6.1140 & 010 -- 000 & 3 1 2 -- 3 0 3 \\
H$_2$O & 6.1166 & 010 -- 000 & 1 1 1 -- 0 0 0 \\
H$_2$O & 6.1434 & 010 -- 000 & 2 0 2 -- 1 1 1 \\
H$_2$O & 6.1596 & 010 -- 000 & 2 1 1 -- 2 0 2 \\
H$_2$O & 6.1857 & 010 -- 000 & 1 1 0 -- 1 0 1 \\
H$_2$O & 6.3447 & 010 -- 000 & 1 0 1 -- 1 1 0 \\
H$_2$O & 6.3705 & 010 -- 000 & 2 0 2 -- 2 1 1 \\
H$_2$O & 6.3740 & 010 -- 000 & 2 2 1 -- 3 1 2 \\
H$_2$O & 6.3905 & 010 -- 000 & 1 1 1 -- 2 0 2 \\
H$_2$O & 6.4094 & 010 -- 000 & 3 1 2 -- 3 2 1 \\
H$_2$O & 6.4165 & 010 -- 000 & 3 0 3 -- 3 1 2 \\
H$_2$O & 6.4205 & 010 -- 000 & 0 0 0 -- 1 1 1 \\
H$_2$O & 6.4338 & 010 -- 000 & 5 1 4 -- 5 2 3 \\
H$_2$    & 6.4388 & 1-- 1  & S( 6) \\
H$_2$O & 6.4848 & 010 -- 000 & 2 1 2 -- 2 2 1 \\
H$_2$O & 6.4925 & 010 -- 000 & 2 1 2 -- 3 0 3 \\
H$_2$O & 6.4978 & 010 -- 000 & 1 0 1 -- 2 1 2 \\
H$_2$O & 6.5195 & 010 -- 000 & 3 2 1 -- 3 3 0 \\
H$_2$O & 6.5226 & 010 -- 000 & 3 1 3 -- 3 2 2 \\
H$_2$O & 6.5423 & 010 -- 000 & 3 2 2 -- 3 3 1 \\
H$_2$O & 6.5554 & 010 -- 000 & 4 2 3 -- 4 3 2 \\
H$_2$O & 6.5676 & 010 -- 000 & 2 0 2 -- 3 1 3 \\
H$_2$O & 6.5738 & 010 -- 000 & 4 1 4 -- 4 2 3 \\
H$_2$O & 6.5904 & 010 -- 000 & 3 1 3 -- 4 0 4 \\
H$_2$O & 6.6287 & 010 -- 000 & 4 2 3 -- 5 1 4 \\
H$_2$O & 6.6421 & 010 -- 000 & 1 1 0 -- 2 2 1 \\
H$_2$O & 6.6723 & 010 -- 000 & 1 1 1 -- 2 2 0 \\
H$_2$O & 6.6836 & 010 -- 000 & 4 1 4 -- 5 0 5 \\
H$_2$O & 6.7078 & 010 -- 000 & 4 0 4 -- 5 1 5 \\
H$_2$O & 6.7123 & 010 -- 000 & 6 1 6 -- 6 2 5 \\
H$_2$O & 6.7235 & 010 -- 000 & 2 1 1 -- 3 2 2 \\
H$_2$O & 6.7745 & 010 -- 000 & 5 1 5 -- 6 0 6 \\
H$_2$O & 6.7865 & 010 -- 000 & 5 0 5 -- 6 1 6 \\
H$_2$O & 6.7934 & 010 -- 000 & 3 1 2 -- 4 2 3 \\
H$_2$O & 6.8265 & 010 -- 000 & 2 1 2 -- 3 2 1 \\
H$_2$O & 6.8531 & 010 -- 000 & 4 1 3 -- 5 2 4 \\
H$_2$O & 6.8577 & 010 -- 000 & 2 2 0 -- 3 3 1 \\
H$_2$O & 6.8642 & 010 -- 000 & 2 2 1 -- 3 3 0 \\
H$_2$ & 6.9098 & 0 -- 0 & S(5) \\
H$_2$O & 6.9600 & 010 -- 000 & 3 2 1 -- 4 3 2 \\
H$_2$O & 6.9934 & 010 -- 000 & 3 2 2 -- 4 3 1 \\
H$_2$O & 7.0242 & 010 -- 000 & 3 1 3 -- 4 2 2 \\
H$_2$O & 7.0451 & 010 -- 000 & 3 3 1 -- 4 4 0 \\
H$_2$O & 7.1182 & 010 -- 000 & 5 2 3 -- 6 3 4 \\
H$_2$O & 7.1471 & 010 -- 000 & 4 2 3 -- 5 3 2 \\
H$_2$O & 7.1647 & 010 -- 000 & 4 3 1 -- 5 4 2 \\
H$_2$O & 7.1714 & 010 -- 000 & 4 3 2 -- 5 4 1 \\
H$_2$O & 7.2079 & 010 -- 000 & 4 4 1 -- 5 5 0 \\
H$_2$O & 7.2726 & 010 -- 000 & 4 1 4 -- 5 2 3 \\
H$_2$ & 7.2802 & 1 -- 1 & S(5) \\
CH$_4$ & 7.4748 & 00011F2 -- 00001A1 & 6F2 2 -- 5F1 1 \\
CH$_4$ & 7.4761 & 00011F2 -- 00001A1 & 6F1 1 -- 5F2 1 \\
CH$_4$ & 7.4791 & 00011F2 -- 00001A1 & 6E 1 -- 6F2 2 \\
CH$_4$ & 7.4797 & 00011F2 -- 00001A1 & 6F2 1 -- 5F1 2 \\
CH$_4$ & 7.5034 & 00011F2 -- 00001A1 & 5A2 1 -- 4A1 1 \\
CH$_4$ & 7.5044 & 00011F2 -- 00001A1 & 5F2 1 -- 4F1 1 \\
CH$_4$ & 7.5051 & 00011F2 -- 00001A1 & 5E 1 -- 4E 1 \\
CH$_4$ & 7.5070 & 00011F2 -- 00001A1 & 5F1 1 -- 4F2 1 \\
CH$_4$ & 7.5335 & 00011F2 -- 00001A1 & 4F2 1 -- 3F1 1 \\
CH$_4$ & 7.5343 & 00011F2 -- 00001A1 & 4F1 1 -- 3F2 1 \\
CH$_4$ & 7.5354 & 00011F2 -- 00001A1 & 4A1 1 -- 3A2 1 \\
CH$_4$ & 7.5634 & 00011F2 -- 00001A1 & 3E 1 -- 2E 1 \\
CH$_4$ & 7.5638 & 00011F2 -- 00001A1 & 3F1 1 -- 2F2 1 \\
CH$_4$ & 7.5940 & 00011F2 -- 00001A1 & 2F2 1 -- 1F1 1 \\
CH$_4$ & 7.6253 & 00011F2 -- 00001A1 & 1A2 1 -- 0A1 1 \\
CH$_4$ & 7.6554 & 00011F2 -- 00001A1 & 6A2 1 -- 6A1 1 \\
CH$_4$ & 7.6561 & 00011F2 -- 00001A1 & 3A1 1 -- 3A2 1 \\
CH$_4$ & 7.6570 & 00011F2 -- 00001A1 & 4F1 2 -- 4F2 1 \\
CH$_4$ & 7.6575 & 00011F2 -- 00001A1 & 5F2 3 -- 5F1 2 \\
CH$_4$ & 7.6581 & 00011F2 -- 00001A1 & 1F2 1 -- 1F1 1 \\
CH$_4$ & 7.6587 & 00011F2 -- 00001A1 & 2F1 1 -- 2F2 1 \\
CH$_4$ & 7.6590 & 00011F2 -- 00001A1 & 6F2 3 -- 6F1 1 \\
CH$_4$ & 7.6595 & 00011F2 -- 00001A1 & 5E 2 -- 5E 1 \\
CH$_4$ & 7.6603 & 00011F2 -- 00001A1 & 2E 1 -- 2E 1 \\
CH$_4$ & 7.6605 & 00011F2 -- 00001A1 & 3F1 2 -- 3F2 1 \\
CH$_4$ & 7.6628 & 00011F2 -- 00001A1 & 6F1 3 -- 6F2 2 \\
CH$_4$ & 7.6637 & 00011F2 -- 00001A1 & 3F2 1 -- 3F1 1 \\
CH$_4$ & 7.6652 & 00011F2 -- 00001A1 & 4E 1 -- 4E 1 \\
CH$_4$ & 7.6673 & 00011F2 -- 00001A1 & 4F2 2 -- 4F1 1 \\
CH$_4$ & 7.6704 & 00011F2 -- 00001A1 & 4A2 1 -- 4A1 1 \\
CH$_4$ & 7.6724 & 00011F2 -- 00001A1 & 5F1 2 -- 5F2 1 \\
CH$_4$ & 7.6759 & 00011F2 -- 00001A1 & 5F2 2 -- 5F1 1 \\
CH$_4$ & 7.6802 & 00011F2 -- 00001A1 & 6A1 1 -- 6A2 1 \\
CH$_4$ & 7.6831 & 00011F2 -- 00001A1 & 6F1 2 -- 6F2 1 \\
CH$_4$ & 7.6842 & 00011F2 -- 00001A1 & 6E 2 -- 6E 1 \\
CH$_4$ & 7.6907 & 00011F2 -- 00001A1 & 0F2 1 -- 1F1 1 \\
CH$_4$ & 7.7239 & 00011F2 -- 00001A1 & 1E 1 -- 2E 1 \\
CH$_4$ & 7.7257 & 00011F2 -- 00001A1 & 1F1 1 -- 2F2 1 \\
CH$_4$ & 7.7582 & 00011F2 -- 00001A1 & 2F2 2 -- 3F1 1 \\
CH$_4$ & 7.7612 & 00011F2 -- 00001A1 & 2F1 2 -- 3F2 1 \\
CH$_4$ & 7.7651 & 00011F2 -- 00001A1 & 2A1 1 -- 3A2 1 \\
CH$_4$ & 7.7914 & 00011F2 -- 00001A1 & 3A2 1 -- 4A1 1 \\
CH$_4$ & 7.7943 & 00011F2 -- 00001A1 & 3F2 2 -- 4F1 1 \\
CH$_4$ & 7.7965 & 00011F2 -- 00001A1 & 3E 2 -- 4E 1 \\
CH$_4$ & 7.8027 & 00011F2 -- 00001A1 & 3F1 3 -- 4F2 1 \\
CH$_4$ & 7.8280 & 00011F2 -- 00001A1 & 4F2 4 -- 5F1 1 \\
CH$_4$ & 7.8318 & 00011F2 -- 00001A1 & 4F1 3 -- 5F2 1 \\
CH$_4$ & 7.8408 & 00011F2 -- 00001A1 & 4E 2 -- 5E 1 \\
CH$_4$ & 7.8429 & 00011F2 -- 00001A1 & 4F2 3 -- 5F1 2 \\
CH$_4$ & 7.8642 & 00011F2 -- 00001A1 & 5E 3 -- 6E 1 \\
CH$_4$ & 7.8653 & 00011F2 -- 00001A1 & 5F1 4 -- 6F2 1 \\
CH$_4$ & 7.8692 & 00011F2 -- 00001A1 & 5A1 1 -- 6A2 1 \\
CH$_4$ & 7.8804 & 00011F2 -- 00001A1 & 5F1 3 -- 6F2 2 \\
CH$_4$ & 7.8841 & 00011F2 -- 00001A1 & 5F2 4 -- 6F1 1 \\
CH$_4$ & 7.8875 & 00011F2 -- 00001A1 & 5A2 2 -- 6A1 1 \\
H$_2$ & 8.0257 & 0 -- 0 & S(4) \\
H$_2$ & 8.4534 & 1 -- 1 & S(4) \\
OH & 9.1337 & X1/2 0  -- X1/2 0   & RR 45.5ee  \\
OH & 9.1384 & X1/2 0  -- X1/2 0 & RR 44.5ee  \\
OH & 9.1546 & X1/2 0  -- X1/2 0 & RR 43.5ff  \\
OH & 9.1813 & X1/2 0 -- X1/2 0         & RR 42.5ff       \\
OH & 9.2186 & X1/2 0 -- X1/2 0 & RR 41.5ff       \\
OH & 9.2664 & X1/2 0 -- X1/2 0        & RR 40.5ff   \\
OH & 9.3251 & X1/2 0  -- X1/2 0   & RR 39.5ff       \\
OH & 9.3947 & X1/2 0 -- X1/2 0  & RR 38.5ff   \\
OH & 9.4757 & X1/2 0 -- X1/2 0 & RR 37.5ff       \\
OH & 9.5686 &  X1/2 0 -- X1/2 0  & RR 36.5ff     \\
H$_2$ & 9.6656 & 0 -- 0 & S(3) \\
OH & 9.6726 & X3/2 0 -- X3/2 0 & RR 36.5ee    \\
OH & 9.6738 & X1/2 0 -- X1/2 0        & RR 35.5ff    \\
OH & 9.7921 & X1/2 0 -- X1/2 0        & RR 34.5ff \\
OH & 9.9241 & X1/2 0 -- X1/2 0    & RR 33.5ff       \\
OH & 10.0707 & X1/2 0 -- X1/2 0    & RR 32.5ff    \\
OH & 10.0707 & X1/2 0 -- X1/2 0    & RR 32.5ff    \\
H$_2$    & 10.1778 & 1--1 & S(3) \\
OH & 10.2329 & X1/2 0 -- X1/2 0    & RR 31.5ff  \\
OH & 10.4120 & X1/2 0 -- X1/2 0      & RR 30.5ff  \\
OH & 10.6093 & X1/2 0 -- X1/2 0 & RR 29.5ff \\
OH & 10.8263 & X1/2 0 -- X1/2 0 & RR 28.5ff \\
OH & 11.0475 & X3/2 0 -- X3/2 0 & RR 28.5ff \\
OH & 11.0522 & X1/2 0 -- X1/2 0 & RR 27.5ee \\
OH & 11.0647 & X1/2 0 -- X1/2 0 & RR 27.5ff \\
OH & 11.3267 & X1/2 0 -- X1/2 0 & RR 26.5ff \\
OH & 11.6103 & X3/2 0 -- X3/2 0 & RR 26.5ee \\
OH & 11.6147 & X1/2 0 -- X1/2 0 & RR 25.5ff \\
OH & 11.6147 & X1/2 0 -- X1/2 0 & RR 25.5ff \\
OH & 11.9314 & X1/2 0 -- X1/2 0 & RR 24.5ff \\
OH & 12.2556 & X3/2 0 -- X3/2 0 & RR 24.5ff \\
H$_2$ & 12.2796 & 0 -- 0 & S(2) \\
OH & 12.6574 & X3/2 0 -- X3/2 0 & RR 23.5ee \\
OH & 12.6646 & X1/2 0 -- X1/2 0 & RR 22.5ff \\
OH & 13.0807 & X3/2 0 -- X3/2 0 & RR 22.5ee \\
OH & 13.0893 & X1/2 0 -- X1/2 0 & RR 21.5ff \\
OH & 13.5492 & X3/2 0 -- X3/2 0 & RR 21.5ee \\
OH & 13.5596 & X1/2 0 -- X1/2 0 & RR 20.5ff \\
HCN & 13.7598 & 0110--0000 & R 4e \\
HCN & 13.8160 & 0110--0000 & R 3e \\
HCN & 13.8726 & 0110--0000 & R 2e \\
HCN & 13.9297 & 0110--0000 & R 1e \\
HCN & 13.9873 & 0110--0000 & R 0e \\
HCN & 14.0295 & 0110--0000 & Q 10e \\
HCN & 14.0324 & 0110--0000 & Q 9e \\
HCN & 14.0350 & 0110--0000 & Q 8e \\
HCN & 14.0373 & 0110--0000 & Q 7e \\
HCN & 14.0393 & 0110--0000 & Q 6e \\
HCN & 14.0410 & 0110--0000 & Q 5e \\
HCN & 14.0425 & 0110--0000 & Q 4e \\
HCN & 14.0436 & 0110--0000 & Q 3e \\
HCN & 14.0445 & 0110--0000 & Q 2e \\
HCN & 14.0451 & 0110--0000 & Q 1e \\
OH & 14.0692 & X3/2 0 -- X3/2 0 & RR 20.5ee \\
OH & 14.0818 & X1/2 0 -- X1/2 0 & RR 19.5ff \\
HCN & 14.1630 & 0110--0000 & P 2e \\
HCN & 14.2225 & 0110--0000 & P 3e \\
HCN & 14.2826 & 0110--0000 & P 4e \\
CO$_2$ & 14.5528 & 01101 -- 00001 & R 24e \\
CO$_2$ & 14.5867 & 01101 -- 00001 & R 22e \\
CO$_2$ & 14.6208 & 01101 -- 00001 & R 20e \\
OH & 14.6482 & X3/2 0 -- X3/2 0 & RR 19.5ee \\
CO$_2$ & 14.6549 & 01101 -- 00001 & R 18e \\
OH & 14.6636 & X1/2 0 -- X1/2 0 & RR 18.5ff \\
CO$_2$ & 14.6892 & 01101 -- 00001 & R 16e \\
CO$_2$ & 14.7235 & 01101 -- 00001 & R 14e \\
CO$_2$ & 14.7580 & 01101 -- 00001 & R 12e \\
CO$_2$ & 14.7925 & 01101 -- 00001 & R 10e \\
CO$_2$ & 14.8271 & 01101 -- 00001 & R 8e \\
CO$_2$ & 14.8618 & 01101 -- 00001 & R 6e \\
CO$_2$ & 14.8966 & 01101 -- 00001 & R 4e \\
CO$_2$ & 14.9315 & 01101 -- 00001 & R 2e \\
CO$_2$ & 14.9564 & 01101 -- 00001 & Q 34e \\
CO$_2$ & 14.9595 & 01101 -- 00001 & Q 32e \\
CO$_2$ & 14.9624 & 01101 -- 00001 & Q 30e \\
CO$_2$ & 14.9652 & 01101 -- 00001 & Q 28e \\
CO$_2$ & 14.9664 & 01101 -- 00001 & R 0e \\
CO$_2$ & 14.9677 & 01101 -- 00001 & Q 26e \\
CO$_2$ & 14.9701 & 01101 -- 00001 & Q 24e \\
CO$_2$ & 14.9722 & 01101 -- 00001 & Q 22e \\
CO$_2$ & 14.9742 & 01101 -- 00001 & Q 20e \\
CO$_2$ & 14.9760 & 01101 -- 00001 & Q 18e \\
CO$_2$ & 14.9777 & 01101 -- 00001 & Q 16e \\
CO$_2$ & 14.9791 & 01101 -- 00001 & Q 14e \\
CO$_2$ & 14.9803 & 01101 -- 00001 & Q 12e \\
CO$_2$ & 14.9814 & 01101 -- 00001 & Q 10e \\
CO$_2$ & 14.9823 & 01101 -- 00001 & Q 8e \\
CO$_2$ & 14.9830 & 01101 -- 00001 & Q 6e \\
CO$_2$ & 14.9835 & 01101 -- 00001 & Q 4e \\
CO$_2$ & 14.9838 & 01101 -- 00001 & Q 2e \\
CO$_2$ & 15.0191 & 01101 -- 00001 & P 2e \\
CO$_2$ & 15.0543 & 01101 -- 00001 & P 4e \\
CO$_2$ & 15.0896 & 01101 -- 00001 & P 6e \\
CO$_2$ & 15.1249 & 01101 -- 00001 & P 8e \\
CO$_2$ & 15.1604 & 01101 -- 00001 & P 10e \\
CO$_2$ & 15.1959 & 01101 -- 00001 & P 12e \\
CO$_2$ & 15.2316 & 01101 -- 00001 & P 14e \\
CO$_2$ & 15.2673 & 01101 -- 00001 & P 16e \\
OH & 15.2674 & X3/2 0 -- X3/2 0 & RR 18.5ff \\
CO$_2$ & 15.3031 & 01101 -- 00001 & P 18e \\
CO$_2$ & 15.3390 & 01101 -- 00001 & P 20e \\
CO$_2$ & 15.3749 & 01101 -- 00001 & P 22e \\
OH & 16.0214 & X3/2 0 -- X3/2 0 & RR 17.5ee \\
OH & 16.0451 & X1/2 0 -- X1/2 0 & RR 16.5ff \\
OH & 16.8404 & X3/2 0 -- X3/2 0 & RR 16.5ee \\
OH & 16.8703 & X1/2 0 -- X1/2 0             & RR 15.5ff \\
H$_2$ & 17.0362 & 0 -- 0 & S(1) \\
OH & 17.7693 & X3/2 0 -- X3/2 0            & RR 15.5ee \\
OH & 17.8076 & X1/2 0 -- X1/2 0           & RR 14.5ff \\
OH & 18.7921 & X3/2 0 -- X3/2 0 & RR 14.5ff \\
OH & 18.8299 & X3/2 0 -- X3/2 0 & RR 14.5ee \\
OH & 18.8487 & X1/2 0 -- X1/2 0 & RR 13.5ee \\
OH & 18.8795 & X1/2 0 -- X1/2 0 & RR 13.5ff \\
OH & 20.0085 & X3/2 0 -- X3/2 0 & RR 13.5ff \\
OH & 20.0497 & X3/2 0 -- X3/2 0 & RR 13.5ee \\
OH & 20.0819 & X1/2 0 -- X1/2 0 & RR 12.5ee \\
OH & 20.1151 & X1/2 0 -- X1/2 0 & RR 12.5ff \\
OH & 21.4199 & X3/2 0 -- X3/2 0 & RR 12.5ff \\
OH & 21.4649 & X3/2 0 -- X3/2 0 & RR 12.5ee \\
OH & 21.5169 & X1/2 0 -- X1/2 0 & RR 11.5ee \\
OH & 21.5525 & X1/2 0 -- X1/2 0 & RR 11.5ff \\
OH & 23.0739 & X3/2 0 -- X3/2 0 & RR 11.5ff \\
OH & 23.1234 & X3/2 0 -- X3/2 0 & RR 11.5ee \\
OH & 23.2048 & X1/2 0 -- X1/2 0 & RR 10.5ee \\
OH & 23.2433 & X1/2 0 -- X1/2 0 & RR 10.5ff \\
OH & 25.0351 & X3/2 0 -- X3/2 0 & RR 10.5ff \\
OH & 25.0899 & X3/2 0 -- X3/2 0 & RR 10.5ee \\
OH & 25.2164 & X1/2 0 -- X1/2 0 & RR 9.5ee \\
OH & 25.2579 & X1/2 0 -- X1/2 0 & RR 9.5ff \\
\enddata
\end{deluxetable}

%% file: Line_List_ion.tex
\begin{deluxetable}{ccc} %[b]
\label{LineListAtomic}
\tablecaption{Detected Atomic and Ionic lines. Line information is taken from NIST \citep{NISThttps://doi.org/10.18434/t4w30f}}
\tablehead{\colhead{Species} & \colhead{Wavelength} & \colhead{Transition} \\ 
\colhead{} & \colhead{(\micron)} & \colhead{} } 

%% All data must appear between the \startdata and \enddata commands
\startdata
     $[$Fe II$]$ & 5.341 & ${}^4 F_{9/2}-{}^6 D_{9/2}$ \\
     $[$Ni II$]$ & 6.635 & ${}^2 D_{3/2}-{}^2 D_{5/2}$ \\
     $[$Fe II$]$ & 6.7235 & ${}^4 F_{9/2}-{}^6 D_{7/2}$ \\
     $[$Ar II$]$ & 6.985 & ${}^2 P_{1/2}-{}^2 P_{3/2}$ \\
     $[$Co II$]$ & 10.523 & ${}^3 F_{3}-{}^3F_{4}$ \\
     $[$Ni II$]$ & 10.682 & ${}^4 F_{7/2}-{}^4 F_{9/2}$ \\
     $[$Cl I$]$ & 11.333 & ${}^2 P_{1/2} - {}^2 P_{3/2}$ \\
     $[$Ni II$]$ & 12.729 & ${}^4 F_{5/2}-{}^4 F_{7/2}$ \\
     $[$Ne II$]$ & 12.814 & ${}^2 P_{1/2} - {}^2 P_{3/2}$ \\
     $[$Fe II$]$ & 17.936 & ${}^4 F_{7/2}-{}^4 F_{9/2}$ \\
     $[$Fe II$]$ & 24.522 & ${}^4 F_{5/2}-{}^4 F_{7/2}$ \\
     $[$S I$]$ & 25.245 & ${}^3 P_1 - {}^3 P_2$ \\
     $[$Fe II$]$ & 25.985 & ${}^6 F_{7/2}-{}^6 F_{9/2}$ \\
\enddata
\end{deluxetable}